\documentclass[reprint,prl,twocolumn,superscriptaddress,showpacs,showkeys,floatfix,preprintnumbers]{revtex4-1}

\usepackage{braket}
\usepackage{xargs}
\usepackage{mathtools}
\usepackage[english]{babel}
\usepackage[utf8]{inputenc}
\usepackage{amsmath,amssymb,braket,slashed,mathrsfs}
\usepackage{pifont}
\usepackage{MnSymbol}
\usepackage{graphicx}
\usepackage{tikz}
\usetikzlibrary{shapes,arrows,positioning}
\tikzset{decision/.style={diamond, draw, fill=blue!20, text width=4.5em, text badly centered, inner sep=0pt}}
\tikzset{block/.style={rectangle, draw, fill=blue!20, text width=10em, text centered, rounded corners, minimum width=3.5cm}}
\tikzset{block1/.style={rectangle, draw, fill=blue!20, text width=18.5em, text centered, rounded corners, minimum width=3.5cm}}
\tikzset{line/.style={draw, -latex, thick}}
\usepackage{color,hyperref}
\usepackage{multirow,array}
\usepackage{physics}
\usepackage{ulem}

\usepackage{cleveref}
\newcommand{\be}{\begin{equation}}
	\newcommand{\ee}{\end{equation}}
\newcommand{\ba}{\begin{eqnarray}}
\newcommand{\ea}{\end{eqnarray}}

\newcommand{\nn}{\nonumber}

\renewcommand{\vec}[1]{\mathbf{#1}}

\newcommand{\innovation}{Collaborative Innovation Center of Quantum Matter, Beijing 100871, China}
\newcommand{\chep}{Center for High Energy Physics, Peking University, Beijing 100871, China}
\newcommand{\pkuphy}{School of Physics, Peking University, Beijing 100871,
	China}

\newcommand{\Uconn}{Department of Physics, University of Connecticut, Storrs, CT 06269, USA}

\newcommand{\KPH}{Institut f\"ur Kernphysik, Johannes Gutenberg-Universit\"{a}t,
        J.J. Becher-Weg 45, 55128 Mainz, Germany}
\newcommand{\PRISMA}{PRISMA$+$ Cluster of Excellence, Johannes Gutenberg-Universit\"{a}t, Mainz, Germany}
\newcommand{\UTK}{ Department of Physics and Astronomy, University of Tennessee, Knoxville, Tennessee 37996, USA}
\newcommand{\SCNT}{Southern Center for Nuclear-Science Theory (SCNT), Institute of Modern Physics, Chinese Academy of Sciences, Guangdong 516000, China}

\begin{document}
	\title{Lattice QCD determination of the $\gamma Z$ box contribution to the proton weak charge}
	
	\author{Zhao-long~Zhang}\affiliation{\pkuphy}
	\author{Xu~Feng}\affiliation{\pkuphy}\affiliation{\innovation}\affiliation{\chep}\affiliation{\SCNT}
	\author{Mikhail~Gorchtein}\affiliation{\KPH}\affiliation{\PRISMA}
	\author{Lu-Chang Jin}\affiliation{\Uconn}
	\author{Chuan Liu}\affiliation{\pkuphy}\affiliation{\innovation}\affiliation{\chep}
	\author{Chien-Yeah~Seng}\affiliation{\UTK}
	
	\date{\today}
	
	\begin{abstract}
We present the first lattice QCD determination of the $\gamma Z$ box contribution to parity-violating electron-proton scattering, $\square_{\gamma Z}$, a key ingredient for the precise tests of the Standard Model via the proton weak charge.
Our calculation covers the electron beam energies up to $E=155$ MeV.
At $E=0$, the vector contribution is known analytically and excluded by convention; for the remaining axial-vector component we obtain $\square_{\gamma Z}(0)=0.00411(9)$.
Incorporating this result implies a value for weak charge of $Q_W^p=0.06987(50)$. The calculated energy dependence of $\square_{\gamma Z}$ is directly relevant to the upcoming P2 experiment at Mainz, which will operate at the optimized beam energy of 155 MeV to extract $Q_W^p$.
Although the systematic error budget in this exploratory study is not yet complete, this work establishes a lattice-QCD framework for computing the $\gamma Z$ box contribution and demonstrates the feasibility of a first-principles approach, providing a foundation for future calculations with improved systematic control.

	\end{abstract}
	
	\maketitle
	
{\bf Introduction} - Indirect searches at the precision frontier provide a powerful probe of physics beyond the Standard Model (BSM). A key example is the determination of the proton weak charge, $Q_W^p$, which quantifies the proton's coupling to the $Z^0$ boson. Because parity violation uniquely characterizes the weak interaction, measuring $Q_W^p$ through parity-violating electron scatterifng (PVES) offers a theoretically clean test of the Standard Model (SM)~\cite{Qweak:2018tjf,SLACE158:2005uay,PREX:2021umo}.

The weak charge is related to the weak mixing angle  $s_W^2(Q^2)=\sin^2\theta_W(Q^2)$ at vanishing momentum transfer. 
At the tree level, $Q_W^{p,\text{LO}} = 1 - 4s_W^2(0)\approx 0.05$
is accidentally suppressed, making it highly sensitive to higher-order SM corrections and possible BSM effects. 
 Including one-loop electroweak radiative corrections (EWRCs) yields~\cite{Marciano:1983ss,Erler:2003yk}
    \begin{equation}
    \label{eq:Qloopdef}
    Q_{W}^{p}=(1+\Delta \rho+\Delta_e)(1-4s_W^2(0)+\Delta_e^{\prime})
    	+\square_{WW}+\square_{ZZ}+\square_{\gamma Z}.
     \end{equation}
Here, $s_W^2(Q^2)$ is precisely measured at the $Z$ pole and evolved to low scales via renormalization group equations.
Since the perturbative evolution fails at small $Q^2$, the extrapolation to $Q^2=0$ relies on dispersive analyses constrained by data~\cite{Czarnecki:2000ic,Erler:2017knj};
recent lattice QCD studies have also begun to provide
direct nonperturbative determinations of the hadronic running~\cite{Ce:2022eix,Conigli:2025qvh}.
The vertex and neutral-current corrections $\Delta \rho$, $\Delta_e$ and $\Delta_e^{\prime}$ are known to high precision~\cite{Erler:2003yk},
and the short-distance box diagrams $\square_{WW}$ and $\square_{ZZ}$ can be evaluated analytically.
In contrast, the $\gamma Z$ box contribution $\square_{\gamma Z}$
involves both short- and long-distance dynamics and thus
require nonperturbative input.

Experimentally, $Q_W^p$ is extracted from the parity-violating asymmetry $A_{PV}$~\cite{Kumar:2013yoa,Qweak:2018tjf}
\begin{equation}\label{eq:Qdef}
\lim_{Q^2\to0}\frac{A_{PV}}{A_0}=Q_W^p-\Delta_{\text{EWRC}}(E),
\end{equation}
where $A_{PV}=(\sigma_{R}-\sigma_{L})/(\sigma_{R}+\sigma_{L})$ is defined in terms of the cross sections for left- and right-handed electron scattering.
The normalization factor $A_0=-\frac{G_F Q^2}{4\sqrt{2}\pi\alpha_{\text{em}}}$ involves the Fermi constant $G_F$ and the fine-structure constant $\alpha_{\text{em}}$.
At tree level, the ratio $A_{PV}/A_0$ is independent of $E$; EWRCs
introduce an energy dependence through $\Delta_{\text{EWRC}}(E)$.
This quantity is defined as the difference between EWRCs evaluated at the experimental beam energy and at zero energy, where the explicit expression at $E=0$ is given in Eq.~(\ref{eq:Qloopdef}). Subtracting
$\Delta_{\text{EWRC}}(E)$ from the measured asymmetry thus allows a direct determination of $Q_W^p$.

Reducing $E$ reduces the size of EWRCs, bringing $A_{PV}/A_{0}$ closer to $Q_W^p$ but at the cost of a reduced event rate, thereby defining the experimental trade-off.
The $Q_{\text{weak}}$ experiment achieved $Q_W^p=0.0719(45)$ at $E=1.16$ GeV~\cite{Qweak:2018tjf}.
Next-generation measurements, including SoLID at Jefferson Lab~\cite{JeffersonLabSoLID:2022iod} and P2 at Mainz~\cite{Becker:2018ggl}, aim for improved sensitivity.
In particular, P2 will use a continuous-wave beam from the MESA accelerator and a solenoidal spectrometer at $E=155$ MeV to reduce the uncertainty to 1.4\%.

Among the EWRCs, the $\square_{\gamma Z}$ correction has long been the dominant source of theoretical uncertainty due to its nonperturbative nature. 
Considerable theoretical efforts~\cite{Sibirtsev:2010zg,Rislow:2010vi,Gorchtein:2011mz,Blunden:2011rd,Rislow:2013vta,Hall:2013hta,Hall:2015loa,Gorchtein:2015qha,Gorchtein:2015naa,Erler:2019rmr,Guo:2023epz,Guo:2023krd} have been devoted to reducing these uncertainties. 
These studies rely on dispersion relations and, in some cases, model assumptions, so systematic uncertainties remain. 
A first-principles lattice QCD determination is therefore highly desirable.

In this work we present the first lattice QCD determination of $\square_{\gamma Z}$, providing (a) $\square_{\gamma Z}(0)$ for an updated determination of $Q_W^p$ 
and (b) $\square_{\gamma Z}(E)$ for $E \le 155$ MeV, where the upper limit corresponds to the P2 operating energy. 
In contrast to previous lattice studies of the $\gamma W$ box in $\beta$ decay~\cite{Feng:2020zdc,Ma:2021azh,Yoo:2023gln,Ma:2023kfr}, the present calculation faces qualitatively new challenges. First, the finite beam energy in PVES admits on-shell intermediate states below the initial $e$–$p$ energy, making analytic continuation of the Euclidean-time integral essential.
Second, the $\gamma Z$ amplitude receives contributions from both vector ($\square_{\gamma Z}^V$) and axial-vector ($\square_{\gamma Z}^A$) hadron parts, involving the $P$-even invariants $T_1$, $T_2$ and the $P$-odd $T_3$, 
rather than only $T_3$ as in $\gamma W$~\cite{Feng:2020zdc}.
Third, a sizable $N\pi$ contribution arises in $\square_{\gamma Z}^V$, requiring a proper treatment of the Euclidean-time truncation and finite-volume effects.

By overcoming these challenges,
we determine the energy dependence of $\square_{\gamma Z}$ and obtain $\square_{\gamma Z}=0.00412(9)$ at $E=0$.
Consequently, the dominant uncertainties in $Q_W^p$ now stem from the input $s_W^2(0)$ in Eq.~(\ref{eq:Qloopdef}) and from the $P$-odd piece in the $\gamma \gamma$ box diagram~\cite{Gorchtein:2016qtl,Erler:2019rmr}.

{\bf Analytic continuation} - For forward PVES $e(l)+p(p)\to e(l)+p(p)$, the hadronic tensor associated with the $\gamma Z$ box diagram is defined as
\begin{equation}\label{eq:tensor}
	\begin{aligned}
		T_{\mu\nu}&=\int d^4x \,e^{iq\cdot x}\bra{p}T\left[J_{\mu}^{em}(x)J_{\nu}^Z(0)\right]\ket{p},
	\end{aligned}
\end{equation}
where $J_{\mu}^{em}$ and $J_{\mu}^Z$ denote the electromagnetic and neutral weak currents, respectively,
\begin{equation}\label{eq:currents}
	J_{\mu}^{\text{em}}=\sum_{f}e_f\overline{\psi}_f\gamma_{\mu}\psi_f,\quad J^{Z}_{\mu}=\sum_{f}\overline{\psi}_f\gamma_{\mu}\left(g_V^{f}-g_A^{f}\gamma_5\right)\psi_f.
\end{equation}
Here $e_f$ is the electric charge of the quark of flavor $f$. $g_V^f=I_{3L}^f-2e_f\sin^2\theta_W$ and $g_A^f=I_{3L}^{f}$ are its vector and axial-vector couplings, with $I_{3L}^f$ denoting the weak isospin of the left-handed component.
The external momenta are $l=(E,\vec{l})$ for the electron and $p=\left(M,\vec{0}\right)$ for the proton, where $E=|\vec{l}|$ and $M$ is the proton mass.
The loop momentum $q=(\nu,\vec{q})$ enters through the Fourier transform in Minkowski spacetime.

A general Lorentz decomposition of $T_{\mu\nu}$ reads
\begin{equation}\label{eq:struc_func_def}
	\begin{aligned}
		T_{\mu\nu}&=\left(-g_{\mu\nu}+\frac{q_{\mu}q_{\nu}}{q^2}\right)T_1(\nu, \vec{q})\\
		&+\frac{1}{p\cdot q}\left(p_{\mu}-\frac{p\cdot q}{q^2}q_{\mu}\right)\left(p_{\nu}-\frac{p\cdot q}{q^2}q_{\nu}\right)T_2\left(\nu, \vec{q}\right)\\
		&-i\epsilon_{\mu\nu\alpha\beta}\frac{q^{\alpha}p^{\beta}}{2p\cdot q}T_3\left(\nu, \vec{q}\right),
	\end{aligned}
\end{equation}
where $T_{1,2}$ correspond to the vector part and $T_3$ to the axial-vector part of $J_\mu^Z$.

The $\gamma Z$ box contribution can then be expressed as
    \begin{equation}\label{eq:gen_box}
    	\square_{\gamma Z}(E)=\alpha_{\text{em}} \int \frac{d^4q}{\left(2\pi\right)^4}\,\sum_{i=1}^{3}V_i(q,l) T_i(\nu, \vec{q}),
    \end{equation}
where the kinematic functions $V_{1\text{–}3}(q,l)$ are collected in the Supplemental Material~\cite{SM}. The hadronic function $T_i(\nu, \vec{q})$ can, in principle,
be obtained from a Laplace transform of the correlation function computed in lattice QCD along the Euclidean time direction.
However, when $M+|\nu|>E_N=\sqrt{M^2+\vec{q}^2}$, the Laplace integral diverges, rendering the transformation ill-defined.

In contrast, the Euclidean correlator admits a well-defined Fourier transform in Euclidean time; the transform yields the Euclidean temporal momentum $\nu_E$, i.e. the variable conjugate to Euclidean time, which is
used as the argument of $T_i(i\nu_E,\vec{q})$.
We therefore deform the contour and analytically continue the integral as
\ba
\square_{\gamma Z}(E)=\square_{\gamma Z}^{\text{Wick}}(E)+\square_{\gamma Z}^{\text{res}}(E)
\ea
where
\ba
\label{eq:Wick_residue}
&&\square_{\gamma Z}^{\text{Wick}}(E)=\alpha_{\text{em}}\int\frac{d^3\vec{q}}{(2\pi)^3}\int_{-i\infty}^{i\infty} \frac{d\nu}{2\pi}\,\sum_{i=1}^{3}V_i(q,l) T_i(\nu, \vec{q})
\nn\\
&&\hspace{0.2cm}=i \alpha_{\text{em}}\int\frac{d^3\vec{q}}{(2\pi)^3}\int_{-\infty}^{\infty} \frac{d\nu_E}{2\pi}
\sum_{i=1}^{3}V_i(i\nu_E,\vec{q},l) T_i(i\nu_E, \vec{q}),
\nn\\
&&\square_{\gamma Z}^{\text{res}}(E)=i\alpha_{\text{em}}\int\frac{d^3\vec{q}}{(2\pi)^3}\operatorname{Res}[V_i(q,l) T_i(\nu, \vec{q})].
\ea
The residue here specifically arises from poles located in the first and third quadrants of the complex $\nu$-plane.
To determine this term, one must examine the pole structure of $V_i(q,l)$ and $T_i(\nu, \vec{q})$. Although $V_i(q,l)$ involves products of the electron, photon, and $Z^0$ propagators (with the electron mass neglected),
\be
V_i(q,l)\propto \frac{i}{(l-q)^2+i\epsilon}\frac{i}{q^2+i\epsilon}\frac{i}{q^2-M_Z^2+i\epsilon},
\ee 
and $T_i(\nu, \vec{q})$ contains hadronic propagators such as
\be
T_i(\nu, \vec{q})\propto \frac{i}{(p-q)^2-M^2+i\epsilon}+\cdots,
\ee
only a single pole from the electron propagator
\be
\label{eq:lepton_pole}
\nu=\nu_e+i\epsilon\equiv E-|\vec{q}-\vec{l}|+i\epsilon,
\ee
contributes to $\square_{\gamma Z}^{\text{res}}(E)$ when $E\ge |\vec{q}-\vec{l}|$.
Therefore,
\be
\operatorname{Res}[V_i(q,l)T_i(\nu,\vec{q})]
=\operatorname{Res}[V_i(q,l)]T_i(\nu_e,\vec{q}).
\ee
For $E>0$, intermediate states can go on shell and the loop amplitude becomes complex. Throughout this work, we retain only the real part of the $\gamma Z$ box contribution.

{\bf Wick contribution} -
To evaluate $T_i(i\nu_E,\vec{q})$, we perform a Fourier transform of the Euclidean hadronic function $\mathcal{H}_{\mu\nu,E}(t_E,\vec{x})=\langle p|T\{J_{\mu,E}^{\text{em}}(t_E,\vec{x})J_{\nu,E}^Z(0)\}|p\rangle$, yielding
\ba
\label{eq:Wick_integral}
&&T_{{\mu\nu},E}(i\nu_E,\vec{q})=\int dt_E\,e^{-i\nu_E t_E}\tilde{\mathcal{H}}_{\mu\nu,E}(t_E,\vec{q}),
\nn\\
&&\tilde{\mathcal{H}}_{\mu\nu,E}(t_E,\vec{q})=\int d^3\vec{x}\,e^{-i\vec{q}\cdot\vec{x}}\mathcal{H}_{\mu\nu,E}(t_E,\vec{x}).
\ea
Here, the subscript $E$ denotes quantities in Euclidean spacetime. The Euclidean and Minkowski currents diff only by a phase
\be
\label{eq:Euclidean_Minkowski}
J_{\mu,E}^{\text{em},Z}(0)=\eta_\mu J_{\mu}^{\text{em},Z}(0),\quad \eta_\mu=\begin{cases}1 & \mu=0 \\ i & \mu=1,2,3\end{cases}.
\ee
This correspondence enables a direct conversion between the Euclidean and Minkowski representations.
Since the conversion is not central to the discussion below, we omit the subscript $E$ hereafter for brevity.

We decompose $T_{{\mu\nu}}(i\nu_E,\vec{q})$ into short- and long-distance parts
\be
T_{{\mu\nu}}(i\nu_E,\vec{q})=T_{{\mu\nu}}^{\text{SD}}(i\nu_E,\vec{q})+T_{{\mu\nu}}^{\text{LD}}(i\nu_E,\vec{q})
\ee
corresponding to the time integrals $\int_{-t_s}^{t_s} dt_E$ and $2\int_{t_s}^\infty dt_E$, respectively. The hadronic input $\mathcal{H}_{\mu\nu}(t_E,\vec{x})$ becomes increasingly noisy at large $t_E$. 
For sufficiently large $t_s$, where ground-state dominance sets in, the long-distance contribution can be reconstructed in two ways:
(i) using the infinite-volume reconstruction (IVR) method~\cite{Feng:2018qpx} with $\mathcal{H}_{\mu\nu}(t_s,\vec{x})$ in coordinate space as input, or (ii) computing the matrix elements $\langle p(\vec{q})|J_\mu^{\text{em},Z}|p\rangle$
from three-point functions, reconstructing $\tilde{\mathcal{H}}_{\mu\nu}(|t_E|\ge t_s,\vec{q})$ and then applying the IVR in the momentum space, as detailed in the Supplemental Material~\cite{SM}. 
The first approach relies solely on four-point correlators and the analysis is simpler, whereas the second also involves three-point correlators but typically achieves higher statistical precision. 
In this work, we adopt method (i) for the axial-vector current and method (ii) for the vector current, whose signal-to-noise ratio is generally poorer.

In Eq.~(\ref{eq:Wick_integral}), the integrals are formally defined in infinite spacetime volume, which is not realized in practical lattice calculations. Nevertheless, the Wick (non-pole) contribution is free from exponentially growing contamination with increasing temporal extent. Moreover, by employing the IVR method, this contribution remains unaffected by power-law finite-volume effects even as 
$\nu_E\to0$.
In contrast, the residue (pole) contribution is more delicate: the exponential growing contamination in the Euclidean-time integral becomes relevant, and multihadron intermediate states can go on shell as the electron energy increases. This feature distinguishes the $\gamma Z$-box contribution in PVES from the $\gamma W$-box contribution in $\beta$ decay.

{\bf Residue contribution} -
Both $\square_{\gamma Z}^V$ and $\square_{\gamma Z}^A$ receive contributions from Wick and residue terms.
In the vector channel, the residue contribution amounts to approximately 30\% of the corresponding Wick term, whereas in the axial channel it is suppressed by roughly an order of magnitude relative to the Wick contribution. Despite its numerical suppression, the evaluation of the residue term is considerably more involved.
Unlike the Wick contribution, the residue contribution requires an input of $T_i(\nu, \vec{q})$ at $\nu=\nu_e$. This requires a Laplace transform of $\mathcal{H}_{\mu\nu}(t_E,\vec{x})$, featuring a different Euclidean time behavior. To make it clear, we explicitly introduce the temporal extent $T$ in the Laplace transform
\be
T_{{\mu\nu}}(\nu_e,\vec{q})=\int_{-T/2}^{T/2} dt_E\,e^{\nu_e t_E}\int d^3\vec{x}\,e^{-i\vec{q}\cdot\vec{x}}\mathcal{H}_{\mu\nu}(t_E,\vec{x}).
\ee
To assess the convergence of the integral, we invoke the spectral representation
\be
T_{\mu\nu}(\nu_e,\vec{q})=\int_{-T/2}^{T/2} dt_E\, e^{\nu_e t_E} \int d\omega \,\rho_{\mu\nu}(\omega,\vec{q})e^{-(\omega-M)|t_E|},
\ee
with $\rho_{\mu\nu}(\omega,\vec{q})$ the spectral weight.
Convergence requires 
\be
\nu_e=E-E_e<E_N-M,
\ee
with $E_e=|\vec{q}-\vec{l}|$ and $E_N=\sqrt{M^2+\vec{q}^2}$ representing the on-shell energies of the intermediate electron and nucleon, respectively. 
For any nonzero beam energy $E$, there exist momenta $\vec{q}$ such that the total energy of the lowest intermediate state, $E_e+E_N$, lies below the initial-state energy $E+M$,
leading the Laplace integral to diverge as $T\to\infty$. This divergence can be overcome by applying the infinite-volume reconstruction method as detailed in the Supplemental Material~\cite{SM}.

When the energy $E$ increases, the location of the pole $\nu_e$ could also lie above the $N\pi$ production threshold, namely
\be
E-E_e>E_{N\pi}-M.
\ee
The threshold beam energy $E_{\text{th}}$ for producing an on-shell 
$N\pi$ state is determined by the kinematic condition
\be
(E_{\text{th}}+M)^2-E_{\text{th}}^2=(M+M_\pi)^2,
\ee
which gives $E_{\mathrm{th}}\simeq150~\mathrm{MeV}$.
Note that the target of this calculation is to compute the box contribution with the range of $0\le E\le 155$ MeV, with the upper limit lying slightly above $E_{\text{th}}$. 
When the energy $E$ is close to $E_{\text{th}}$, whether approached from above or below, power-law finite-volume effects must be taken into account. In addition, the long-distance reconstruction receives contributions from low-lying $N\pi$ states. See the Supplemental Material~\cite{SM} for further details.

{\bf Numerical Results} - 
In this work, we use the 24D and 32Dfine gauge ensembles at near-physical pion mass, generated by the RBC–UKQCD Collaboration with 2+1-flavor domain-wall fermions~\cite{RBC:2014ntl}.
The ensembles have pion masses of 142.6(3) MeV (24D) and 142.9(7) MeV (32Dfine), similar spatial volumes $L = 4.6$ fm, and lattice spacings $a^{-1} = 1.023(2)$ GeV and 1.378(5) GeV, respectively.
With two lattice spacings and one lattice volume, it is not sufficient for us to fully assess higher-order discretization effects and to perform a systematic study of finite-volume effects. Regarding the latter,
we have performed a detailed three-step analysis in the Supplementary Material~\cite{SM}, which indicates that the dominant power-law finite-volume effects expected in this calculation are consistent with being small at the present volume, and that no significant residual effects are resolved within the current statistical precision.

To determine the Compton amplitude, we compute the nucleon four-point function
	\begin{equation}\label{eq:4ptfunc}
		\sum _{\vec{x}_f,\vec{x}_i}\operatorname{Tr}\left[\mathscr{P}\langle O_{p}(t_f,\mathbf{x}_f)J_\mu^{em}(x)J_\nu^{Z}(y)\bar{O}_{p}(t_i,\mathbf{x}_i)\rangle\right],
	\end{equation}
using the random-sparse-field method~\cite{Li:2020hbj,Detmold:2019fbk}.
Here $\mathscr{P}$ denotes the parity projector, and $O_{p}$ ($\bar{O}_{p}$) are the proton annihilation (creation) operators. Smeared sources and sinks are employed to improve the signal.
Local vector
and axial vector current operators are contracted with
the renormalization factors quoted from Ref.~\cite{Feng:2021zek}.

We comment here on the treatment of quark-line-disconnected contributions. Such contributions enter in two distinct ways. First, in the $\gamma Z$ box contribution, the electromagnetic and weak currents can form separate quark loops. These disconnected contractions are neglected, as they are suppressed by flavor $SU(3)$ symmetry. Second, disconnected contributions can enter the reconstruction of the relevant intermediate states. These contributions are also neglected as they are expected to be suppressed by the Okubo-Zweig-Iizuka rule. A quantitative assessment of these contributions is left for future work.

The time-slice convention is chosen as
 \begin{equation}
 	t_i=\min\left\{t_x,t_y\right\}-\Delta t_i,\quad t_f=\max\left\{t_x,t_y\right\}+\Delta t_f,
 \end{equation}
so that $|t_x-t_y|$ represents the current-current separation, while $\Delta t_i+\Delta t_f$ controls the excited-state contamination from the nucleon interpolators.

The vector $\gamma Z$ box, $\square_{\gamma Z}^{V}(E)$, corresponds to the Compton amplitude in which both currents are vector. 
As indicated in Eq.~(\ref{eq:gen_box}), $\square_{\gamma Z}^{V}(E)$ depends solely on the invariant amplitudes $T_{1,2}(q)$,
 \begin{equation}
 \label{eq:vectorgZ}
 	\square_{\gamma Z}^{V}(E)=\alpha_{\text{em}} \int \frac{d^4 q}{\left(2\pi\right)^4}\sum_{i=1}^{2}V_i(q,l)T_i(\nu,\vec{q}).
 \end{equation}
To isolate the short-distance (SD) part of the Compton amplitude, we restrict the Euclidean time separation to $|t_E|<t_s$ and define
 \be
 \label{eq:T_SD}
T_{\mu\nu}^{\text{SD}}(\nu,\vec{q};t_s)=\int_{-t_s}^{t_s}dt_E\int d^3\vec{x}\,e^{\nu t_E}e^{-i\vec{q}\cdot\vec{x}}\,\mathcal{H}_{\mu\nu}(t_E,\vec{x})
 \ee
The long-distance (LD) contribution is defined by the complementary region $|t_E|>t_s$; for sufficiently large $t_s$ it is dominated by the lowest intermediate states, such as $N$ and $N\pi$.
Compared with $\square_{\gamma Z}^{A}(E)$, the vector contribution is substantially noisier and therefore requires advanced noise-reduction techniques; further details are given in the Supplemental Material~\cite{SM}.

In Fig.~\ref{fig:vector}, we present the lattice results for both the Wick and residue contributions to $\square_{\gamma Z}^{V}$, separated into the short-distance (SD) part with $|t|<t_s$ and the long-distance (LD) part with $|t|>t_s$. The latter receives the contributions from intermediate nucleon and $N\pi$ states, shown separately and labeled as ``$N,\mathrm{LD}$'' and ``$N\pi,\mathrm{LD}$'', respectively. For the Wick contribution, the $N\pi$ contribution is negligible, whereas in the residue contribution it becomes dominant even at moderately long distances, e.g. $t_s=1.15\ \mathrm{fm}$. This highlights the essential role of reconstructing the $N\pi$ intermediate-state contribution at long distances. Details are given in the Supplemental Material~\cite{SM}.

\begin{figure}[htbp]
	\centering
	\includegraphics[scale=0.45]{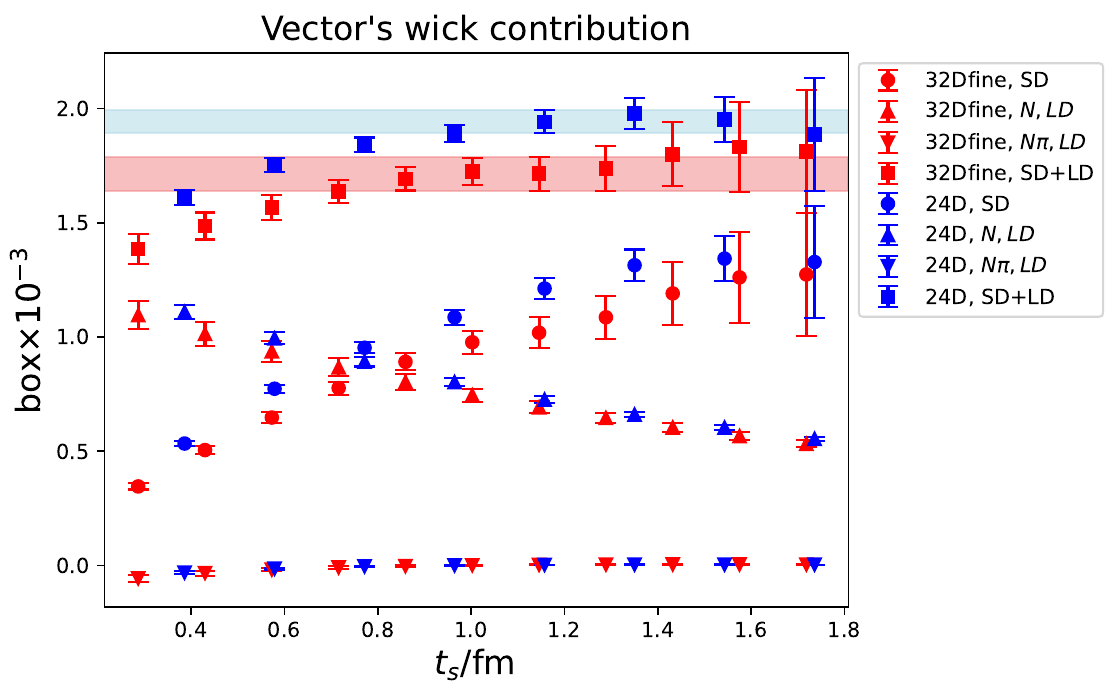}	
	\includegraphics[scale=0.45]{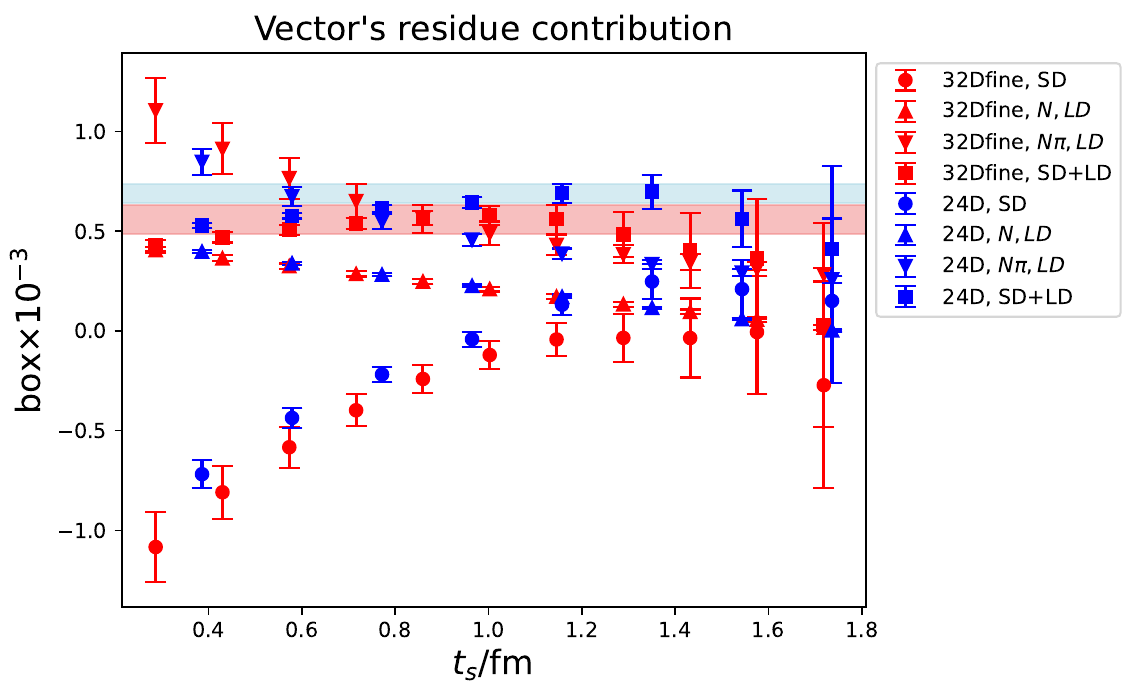}
	\caption{Numerical results for the Wick and residue contributions to $\square_{\gamma Z}^{V}(E)$ at a beam energy of $E=155$ MeV. The short-distance part, the long-distance nucleon contribution, the long-distance $N\pi$ contribution, and their total are indicated by circles, upward triangles, downward triangles, and squares, respectively.}
	\label{fig:vector}
\end{figure}

In the axial $\gamma Z$ box, only $T_3(q)$ in Eq.~(\ref{eq:gen_box}) contributes,
 \begin{equation}\label{eq:axialgZ}
	\square_{\gamma Z}^{A}(E)=\alpha_{\text{em}} \int \frac{d^4 q}{\left(2\pi\right)^4}V_3(q,l)T_3(\nu,\vec{q}).
\end{equation}
As shown in the Supplemental Material~\cite{SM}, the loop integral in Eq.~(\ref{eq:axialgZ}) is ultraviolet finite when the full $Z$-boson propagator is retained. However, replacing $1/(Q^2+M_Z^2)$ with $1/M_Z^2$, with $Q^2=-q^2$ spacelike momentum, makes the integral ultraviolet divergent, indicating a strong sensitivity to the large-$Q^2$ region. In lattice QCD, the momentum is effectively cut off at $\pi/a$, introducing artifacts that scale as $a^2 Q^2$. This differs from the vector box, where the integrand is further suppressed at large $Q^2$ by a factor $\Lambda_{\mathrm{QCD}}^2/Q^2$, yielding lattice artifacts of order $a^2 \Lambda_{\mathrm{QCD}}^2$ even when $Q^2 \simeq M_Z^2$.

Following Refs.~\cite{Feng:2020zdc,Ma:2023kfr}, we split the momentum integral into a nonperturbative region $Q^2<Q_{\mathrm{cut}}^2$ and a perturbative region $Q^2>Q_{\mathrm{cut}}^2$, adopting $Q_{\mathrm{cut}}^2=2~\mathrm{GeV}^2$ as in Ref.~\cite{Erler:2019rmr}. The axial $\gamma Z$ box can then be written as
\begin{equation}\label{eq:axialgZ_split}
	\square_{\gamma Z}^{A}(E)=\square_{\gamma Z}^{A,Q_{\text{cut}}^2<2\text{GeV}^2}(E)+\square_{\gamma Z}^{A,Q_{\text{cut}}^2>2\text{GeV}^2}(E).
\end{equation}
where the low-$Q^2$ contribution is computed nonperturbatively on the lattice, while the high-$Q^2$ contribution is determined using perturbative QCD.
Details are summarized in the Supplemental Material~\cite{SM}.

In Fig.~\ref{Fig:E_dep_all}, we show the vector and axial-vector $\gamma Z$ box as functions of the beam energy $E$ for the 24D and 32Dfine ensembles, together with the continuum-extrapolated results.
These results can be compared with phenomenological determinations based on inclusive scattering data for the vector~\cite{Gorchtein:2015naa} and axial-vector~\cite{Erler:2019rmr} contributions.
(For $\square_{\gamma Z}^{V}$, Ref.~\cite{Gorchtein:2015naa} includes only the inelastic contribution; for comparison, we therefore add the corresponding Born term obtained from phenomenological analyzes.)
After continuum extrapolation, the lattice results lie systematically below the phenomenological estimates for both cases. The values of the $\gamma Z$ box at $E=0$ and $E=155$ MeV are summarized in Table~\ref{tab:combined_cont}.
At $E=0$, the vector $\gamma Z$ box is known analytically, namely
$\square_{\gamma Z}^V(0)=\pi\alpha_{\text{em}} Q_W^{p,\text{LO}}$; a derivation is provided in the Supplemental Material~\cite{SM}.

\begin{figure}[htbp] 
	\centering
	\includegraphics[scale=0.3]{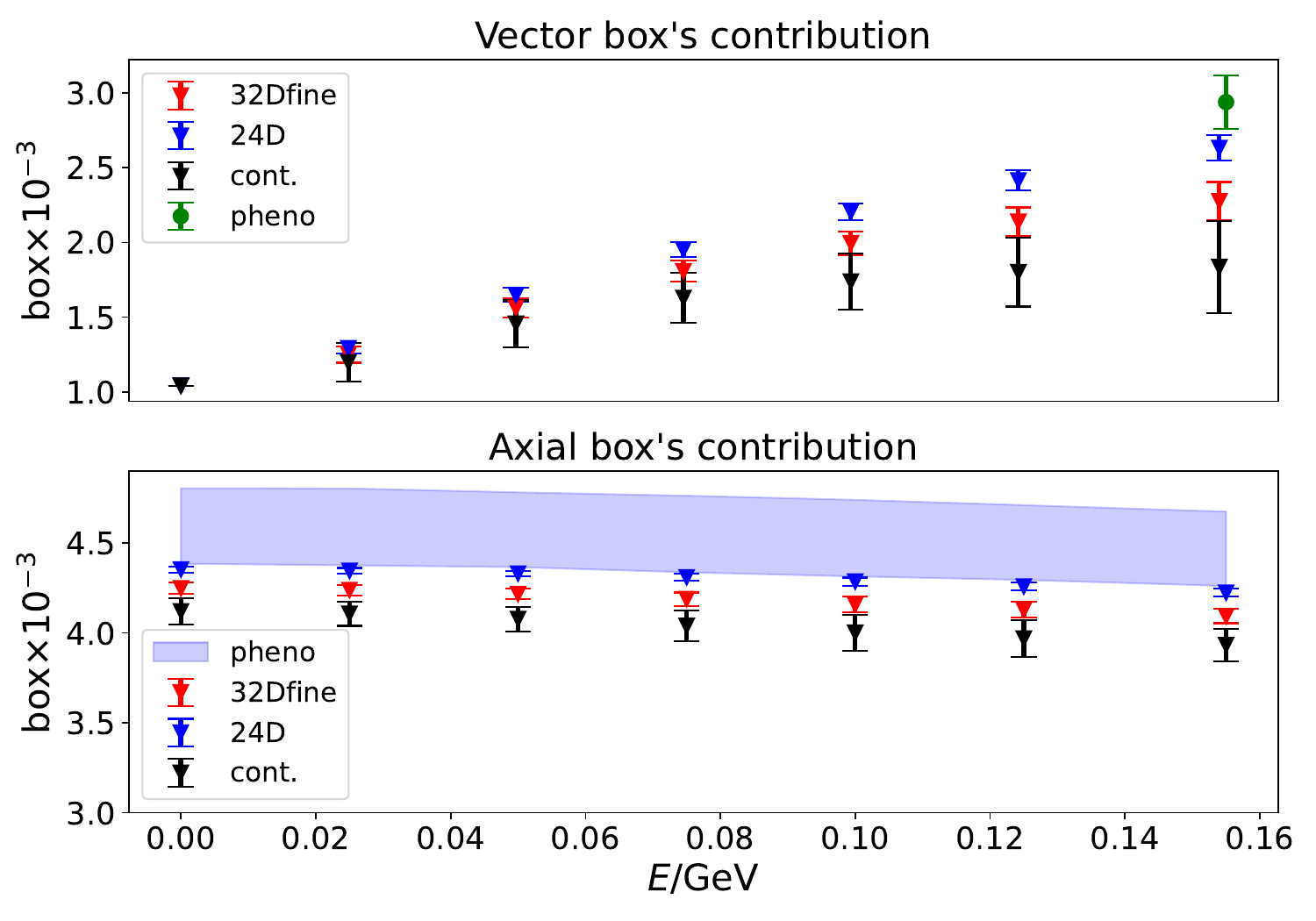}
	\caption{The beam energy dependence of vector box $\square_{\gamma Z}^{V}(E)$ and axial-vector box $\square_{\gamma Z}^{A}(E)$, with 24D, 32Dfine and continuum-extrapolated result.}
	\label{Fig:E_dep_all}
\end{figure}

\begin{table}[htbp]
	\centering
	\begin{tabular}{ccccc}
		\hline
		\hline
& 32Dfine & 24D & Cont. & Pheno \\ 
 \hline 
$\square_{\gamma Z}^V(0\text{ MeV})$& $1.041$ & $1.041$ & $1.041$ & $1.041$ \\ 
$\square_{\gamma Z}^A(0\text{ MeV})$& $4.243(31)$ & $4.347(17)$ & $4.114(72)$ & $4.60(21)$ \\ 
$\square_{\gamma Z}(0\text{ MeV})$& $5.284(31)$ & $5.388(17)$ & $5.156(72)$ & $5.64(21)$ \\ 
\hline
\hline
$\square_{\gamma Z}^V(155\text{ MeV})$& $2.28(13)$ & $2.633(87)$ & $1.84(31)$ &
$2.94(18)$\\ 
$\square_{\gamma Z}^A(155\text{ MeV})$& $4.087(39)$ & $4.218(20)$ & $3.927(91)$ & $4.46(21)$ \\ 
$\square_{\gamma Z}(155\text{ MeV})$& $6.36(13)$ & $6.851(91)$ & $5.76(32)$ & $7.39(28)$ \\ 
\hline
	\end{tabular}%
	\caption{Numerical results of $\gamma Z$ box at beam energy $E=0$ and 155 MeV. All the results are given in unit $10^{-3}$.}
	\label{tab:combined_cont}
\end{table}

All results shown in Fig.~\ref{Fig:E_dep_all} are obtained from lattice data with $\Delta t_i+\Delta t_f\simeq0.75$ fm. As documented in the Supplemental Material~\cite{SM}, we observe clear plateaus in the range
$0.7~\text{fm}\lesssim \Delta t_i+\Delta t_f\lesssim1.2~\text{fm}$ for both vector and axial-vector contributions on both ensembles. We therefore quote results at $\Delta t_i+\Delta t_f\simeq0.75$ fm as our central values. Residual excited-state effects are estimated by comparing with a two-state fit and are included as a systematic uncertainty.
At $E=155$ MeV, finite-volume effects arising from intermediate $N\pi$ states are assessed using the three-step procedure detailed in the Supplemental Material~\cite{SM}. We find that leading-order finite-volume corrections contribute only 3-5\% to the vector $\gamma Z$ box, corresponding to a 1-2\% effect in the total $\gamma Z$ box. At $E=0$, where the $N\pi$ threshold is closed, such effects can be safely neglected.
Our final results are
\ba
&&\square_{\gamma Z}(0)=5.16(7)_{\mathrm{stat}}(2)_{a}(5)_{\mathrm{ES}}\times10^{-3},
\nn\\
&&\square_{\gamma Z}(155\text{ MeV})=5.76(32)_{\mathrm{stat}}(9)_{a}(26)_{\mathrm{ES}}(11)_{\mathrm{FV}}\times10^{-3},
\nn\\
\ea
where the uncertainties are statistical (stat), residual lattice discretization effect ($a$), excited-state contamination (ES), and finite-volume effects (FV), respectively. The continuum extrapolation is based on two lattice spacings. We estimate the residual discretization uncertainty associated with higher-order terms using Symanzik power counting, taking the coefficient of the $a^4$ term to be of natural size relative to that of the leading $a^2$ term. Specifically, we replace the linear fit, $c_0+c_1 a^2\Lambda_{\rm QCD}^2$, by $c_0+c_1 a^2\Lambda_{\rm QCD}^2\pm c_1 a^4\Lambda_{\rm QCD}^4$ and use the resulting variation as a rough estimate of the residual effect. We emphasize that this procedure provides an estimate of the expected size of higher-order discretization effects, rather than a data-driven determination of the continuum-extrapolation systematic uncertainty.
Calculations at finer lattice spacings will be required to fully quantify residual discretization effects.

In relating $Q_W^p$ to $s_W^2(0)$ in Eq.~(\ref{eq:Qloopdef}), 
earlier analyses - including Qweak experiment~\cite{Qweak:2018tjf} and dispersive studies~\cite{Hall:2015loa} - explicitly excluded the vector $\gamma Z$ box at $E=0$
to avoid double counting, following the classic atomic-physics treatment~\cite{Marciano:1983ss}. Adopting the same convention and using the subscript ``sub'', we take
\be
\square_{\gamma Z,\rm{sub}}(0)=\square_{\gamma Z}^A(0)=4.11(7)_{\mathrm{stat}}(2)_{a}(5)_{\mathrm{ES}}\times10^{-3},
\ee
which yields
\be
Q_W^{p}=0.06987(20)_{s_{W}^2}(10)_{WW}(9)_{\gamma Z}(44)_{\gamma\gamma}.
\ee
Here we use $s_W^2(0)=0.23873(5)$, $\Delta_\rho=6.0\times10^{-4}$~\cite{ParticleDataGroup:2024cfk}, $\Delta_e=-0.001161$, $\Delta_e'=-0.001411$,
$\square_{ZZ}=0.00185$, and $\square_{WW}=0.0183(1)$~\cite{Erler:2004in}.
Only uncertainties exceeding $10^{-5}$ are explicitly shown.
In addition, we include the two-loop correction $\square_{\gamma\gamma}^{\mathrm{PV}}=-0.26(44)\times10^{-3}$ taken from Ref.~\cite{Erler:2019rmr}, which is not captured by the one-loop expression in Eq.~(\ref{eq:Qloopdef}).

{\bf Conclusion} - We have presented a first-principles lattice QCD calculation of the $\gamma Z$ box contribution to PVES, with particular emphasis on the complete treatment of intermediate $N\pi$ states in the vector channel. A novel framework is introduced that cleanly separates the box contribution into Wick and residue components. 
We provide the nonperturbative results at both $E=0$ and $E=155$ MeV, the beam energy relevant for the upcoming P2 experiment at Mainz.
This work demonstrates a viable path toward a high-precision lattice determination of the $\gamma Z$ box contribution. Full systematic control in the future
will require additional gauge ensembles at finer lattice spacings and larger volumes.

\begin{acknowledgements}

We gratefully acknowledge many helpful discussions with our colleagues from the
RBC-UKQCD Collaborations.
We thank Peter Blunden and Ross Young for useful communications.
X.F., C.L. and Z.L.Z. were supported in part by NSFC of China under Grants 
No. 12125501, No. 12550007, No. 125B2104, No. 12293060 and No. 12293063.
L.C.J. acknowledges support by DOE Office of Science Early Career Award DE-SC0021147, DOE grant DE-SC0010339 and DE-SC0026314.
The work of M.G. is supported in part by EU Horizon 2020 research and innovation programme, STRONG-2020 project
under grant agreement No 824093, and by the Deutsche Forschungsgemeinschaft (DFG) under the grant agreement GO 2604/3-1.
The work of C.-Y.S. is supported in
part by the DOE Topical Collaboration "Nuclear Theory for New Physics", award No. DE-SC0023663, and by University of Tennessee, Knoxville.
The research reported in this work was carried out using the computing facilities at Chinese National Supercomputer Center in Tianjin.
It also made use of computing and long-term storage facilities of the USQCD Collaboration, which are funded by the Office of Science of the U.S. Department of Energy.

\end{acknowledgements}

	\bibliography{./ref.bib}

\clearpage

\setcounter{page}{1}
\renewcommand{\thepage}{Supplementary Information -- S\arabic{page}}
\setcounter{table}{0}
\renewcommand{\thetable}{S\,\Roman{table}}
\setcounter{equation}{0}
\renewcommand{\theequation}{S\,\arabic{equation}}
\setcounter{figure}{0}
\renewcommand{\thefigure}{S\,\arabic{figure}}

\section{Supplementary Information}

\subsection{Explicit form of weight functions $V_{1\sim 3}$}
The $\gamma Z$ box contribution can be written as
    \begin{equation}
    \label{eq:gamma_Z_box}
    	\square_{\gamma Z}(E)=\alpha_{\text{em}} \int \frac{d^4q}{\left(2\pi\right)^4}\,\sum_{i=1}^{3}V_i(q,l) T_i(\nu, \vec{q}).
    \end{equation}
where $q$ and $l$ denote the photon and lepton four-momenta.
The weight functions $V_{1\sim 3}(q,l)$ are
\ba
\label{eq:Vi_def}
		V_1(q,l)&=&Cg_A^e\frac{l\cdot q}{q^2}\left(-2l\cdot q+3q^2\right)
		\nn\\
		V_2\left(q,l\right)&=&-\frac{Cg_A^{e}}{\left(p\cdot q\right)(q^2)^2}\left[2\left(l\cdot q\right)^2\left(p\cdot q\right)^2+2\left(l\cdot p\right)^2\left(q^2\right)^2\right.
		\nn\\
		&&\left.+\left(l\cdot q\right) q^2\left(-4\left(l\cdot p\right)\left(p\cdot q\right)-\left(p\cdot q\right)^2+M^2q^2\right)\right]
		\nn\\
		V_3\left(q,l\right)&=&Cg_V^e\left(-l\cdot q+\frac{l\cdot p}{p\cdot q}q^2\right),
\ea
with $g_V^e=-\frac{1}{2}+2\sin^2\theta_W$, $g_A^e=-\frac{1}{2}$ and
\ba
C=\frac{16\pi}{l\cdot p}\frac{M_Z^2}{\left(\left(l-q\right)^2+i\epsilon\right)\left(q^2+i\epsilon\right)\left(M_Z^2-q^2-i\epsilon\right)}.
\ea
We decompose the $\gamma Z$ box into Wick and residue contributions with
	\ba
	\label{eq:res&wick}
			\square_{\gamma Z}^{\text{Wick}}(E)&=&\alpha_{\text{em}} \int \frac{d^4Q}{\left(2\pi\right)^4}\sum_{i=1}^{3}V^{\text{Wick}}_i(\nu_E, \vec{q},l) T_i(i\nu_E,\vec{q})
			\nn\\
			\square_{\gamma Z}^\text{res}(E)&=&\alpha_{\text{em}} \int \frac{d^3\vec{q}}{\left(2\pi\right)^3}\sum_{i=1}^{3}V_i^{\text{res}}(\vec{q},l) T_i(\nu_e, \vec{q})
	\ea
with $Q=(\nu_E,\vec{q})$ the Euclidean momentum. The Wick- and residue-weight functions are defined by
\ba
&&V^{\text{Wick}}_i(\nu_E, \vec{q},l)=iV_i(i\nu_E,\vec{q},l),
\nn\\
&&V_i^{\text{res}}(\vec{q},l)=i\operatorname{Res}[V_i(q,l)]=i\lim_{\nu\to\nu_e} (\nu-\nu_e)V_i(q,l).
\nn\\
\ea
Because $T_i(\nu,\vec{q})$ depends only on $\vec{q}^{2}$ (SO(3) invariant) and the proton is at rest, one can perform a solid-angle average over the lepton direction.
After this averaging, the Wick weights take the form
\begin{equation}\label{eq:wick_weight}
	\begin{aligned}
		V_1^{\text{Wick}}&=C^{\text{wick}}ig_A^e\left[\frac{2}{Q^2}+\frac{1}{4E|\vec{q}|}A(l,q)\right]\\
		V_2^{\text{Wick}}&=-C^{\text{wick}}Mg_A^e\left[\frac{1}{Q^2\nu_E} +\frac{1}{|\vec{q}|Q^2}B(l,q)\right.\\
		&\quad \left.+\left(\frac{1}{8E|\vec{q}|\nu_E}-\frac{E}{2|\vec{q}|Q^2\nu_E}\right)A(l,q)\right]\\
		V_3^{\text{Wick}}&=-\frac{1}{4}C^{\text{wick}}g_V^e\left[\frac{1}{E|\vec{q}|}B(l,q)-\frac{1}{|\vec{q}|\nu_E}A(l,q)\right],
	\end{aligned}
\end{equation}
with
\ba
		C^{\text{Wick}}&=&\frac{8\pi}{ME}\frac{m_Z^2}{m_Z^2+Q^2}
		\nn\\
		A(l,q)&=&\ln\left| \frac{Q^2+4E^2-4E|\vec{q}|}{Q^2+4E^2+4E|\vec{q}|}\right| 
		\nn\\
		B(l,q)&=&\arctan\frac{Q^2-2E|\vec{q}|}{2E\nu_E}-\arctan\frac{Q^2+2E|\vec{q}|}{2E\nu_E}.
\ea
The residue contribution depends explicitly on the pole location $\nu=\nu_e=E-|\vec{q}-\vec{l}|$.  Requiring $\nu_e>0$ imposes a constraint on the angle $\theta$ between $\vec{l}$ and $\vec{q}$,
so we retain the explicit $\theta$ dependence in $V_{1\sim 3}^{\text{res}}$
\ba
\label{eq:residue_weight}
		V_1^{\text{res}}&=&-2iC^{\text{res}}Eg_A^{e}
		\nn\\
		V_2^{\text{res}}&=&-iC^{\text{res}}g_A^e\frac{ME\left(\nu_e+|\vec{q}|\cos\theta-2E\right)}{\nu_e\left(\nu_e-|\vec{q}|\cos\theta\right)}
		\nn\\
		V_3^{\text{res}}&=&-iC^{\text{res}}g_V^eE\left(\frac{2E}{\nu_e}-1\right)
\ea
with
\be
		C^{\text{res}}=\frac{4\pi}{ME^2\left(E-\nu_e\right)}\theta_H(2E-|\vec{q}|)\theta_H\left(\cos\theta-\frac{|\vec{q}|}{2E}\right).
\ee
Here $\theta_H$ is a Heaviside step function.

\subsection{High-loop-momentum contributions to $\gamma Z$ box}

At large momentum transfer $Q^2$ with $Q^2=-\nu^2+\vec{q}^2$, the scalar amplitudes $T_i(\nu,\vec{q})$ admit an operator product expansion (OPE)
\be
T_i(\nu,\vec{q})=\sum_nC_{i,n}\left(\frac{Q^2}{\mu^2},\alpha_s\right)\frac{\nu^n}{(Q^2)^n}A_{i,n}(\mu)+\text{H.T.},
\ee
where $A_{i,n}=\langle p|O_{i,n}|p\rangle$ denotes the nucleon matrix element of the local operator $O_{i,n}$. H.T. indicates the higher-twist contributions.
Equivalently, the amplitudes may be regarded as functions of the two invariants $(\nu,Q^2)$. Crossing symmetry implies
\ba
&&T_1(-\nu,Q^2)=T_1(\nu,Q^2),
\nn\\
&&T_2(-\nu,Q^2)=-T_2(\nu,Q^2),
\nn\\
&&T_3(-\nu,Q^2)=-T_3(\nu,Q^2).
\ea
As a consequence, the OPE contains only even or odd powers of $\nu$
\ba
T_1(\nu,\vec{q})&=&\sum_{n=1}C_{1,2n}\left(\frac{Q^2}{\mu^2},\alpha_s\right)\frac{\nu^{2n}}{(Q^2)^{2n}}A_{1,2n}+\text{H.T.}
\nn\\
T_2(\nu,\vec{q})&=&\sum_{n=0}C_{2,2n}\left(\frac{Q^2}{\mu^2},\alpha_s\right)\frac{\nu^{2n+1}}{(Q^2)^{2n+1}}A_{2,2n+1}+\text{H.T.}
\nn\\
T_3(\nu,\vec{q})&=&\sum_{n=0}C_{3,2n}\left(\frac{Q^2}{\mu^2},\alpha_s\right)\frac{\nu^{2n+1}}{(Q^2)^{2n+1}}A_{3,2n+1}+\text{H.T.}
\nn\\
\ea
Inserting the OPE into Eq.~(\ref{eq:gamma_Z_box}), one finds that the integrands associated with $T_{1,2}$ scale at large $Q^2$ as
$\frac{1}{Q^6}\frac{M_Z^2}{M_Z^2+Q^2}$. 
Consequently, these contributions remain ultraviolet convergent even in the absence of the $Z$-boson propagator. 
In contrast, the $T_3$ contribution behaves as $\frac{1}{Q^4}\frac{M_Z^2}{M_Z^2+Q^2}$ at large $Q^2$, which generates a logarithmic enhancement
$\sim\ln\frac{M_Z^2}{\Lambda_{\text{QCD}}^2}$.
As a result, the $T_3$ term is sensitive to the ultraviolet region and requires a matching procedure in which lattice results at $Q^2<Q_{\text{cut}}^2$
are combined with perturbative QCD calculations for $Q^2>Q_{\text{cut}}^2$.

\subsection{Vector $\gamma Z$ box at $E=0$}
In the limit $E\to 0$, the residue contribution vanishes due to $\theta_H(2E-|\vec{q}|)$ factor in Eq.~(\ref{eq:residue_weight}). 
Consequently, only the Wick contribution survives, which can be written as
\ba
\label{eq:wickall}
        \square_{\gamma Z}^{V,\text{Wick}}(E)&=&\frac{\alpha_{\text{em}}}{4\pi^3} \int dQ\,d\nu_E
        \,Q\sqrt{Q^2-\nu_E^2}
	\nn\\
        &&\sum_{i=1}^{2}V^{\text{Wick}}_i(\nu_E, \vec{q},E) T_i(i\nu_E,\vec{q}).
\ea
We separate the integral into two regions, $Q>\epsilon$ and $Q<\epsilon$. For $Q>\epsilon$, the Wick-weight functions admit the expansions
\ba
&&\frac{2}{Q^2}+\frac{1}{4E|\vec{q}|}A(l,q)=\frac{8(3Q^2-4|\vec{q}|^2)}{3Q^6}E^2+\mathcal{O}(E^4),
\nn\\
&&B(l,q)=-\frac{8|\vec{q}|\nu_E}{Q^4}E^2+\mathcal{O}(E^4),
\ea
which imply $V_{1,2}^{\text{wick}}\propto E$ as $E\to0$. Hence, the contributions from the region $Q>\epsilon$ vanish in this limit.

For $Q<\epsilon$, the behavior of $T_{1,2}$ near $Q\simeq 0$ becomes relevant. Gauge invariance constrains the inelastic components to satisfy
\be
T_1^{\text{inel}}(0,\vec{0})=i\frac{3}{2}Q_W^{p,\text{LO}},\quad T_2^{\text{inel}}(0,\vec{0})=0,
\ee
from which it follows that the inelastic contribution to $\square_{\gamma Z}^{V,\text{Wick},Q<\epsilon}(E=0)$ vanishes in the limit $\epsilon\to0$.

The elastic contribution is governed by the Born term, defined as
\begin{equation}
	\begin{aligned}
		T_{\mu\nu}^{\text{Born}}&=\frac{i}{\left(p+q\right)^2-M^2}\bra{p}J_{\mu}^{\text{em}}\ket{p(q)}\bra{p(q)}J_{\nu}^{Z}\ket{p} \\
		&\quad\quad +\{J^{em}_{\mu}\leftrightarrow J_{\nu}^{Z}, q\leftrightarrow -q\},
	\end{aligned}
\end{equation}
with the current matrix elements parameterized as
\ba
\label{eq:gZSach_def}
&&\bra{p(p_f)}J_{\mu}^{\text{em}}\ket{p(p_i)}
\nn\\
&&\quad=\overline{u}(p_f)\left(\gamma_{\mu}F_1^{p}(Q^2)+\frac{i\sigma_{\mu\nu}q^{\nu}}{2M}F_2^{p}(Q^2)\right)u(p_i)
\nn\\		
&&\bra{p(p_f)}J_{\mu}^{Z}\ket{p(p_i)}
\nn\\
&&\quad=\overline{u}(p_f)\left(\gamma_{\mu}F_1^{Z}(Q^2)+\frac{i\sigma_{\mu\nu}q^{\nu}}{2M}F_2^{Z}(Q^2)\right)u(p_i),
\nn\\
\ea
where $q=p_f-p_i$ and $Q^2=-q^2$.
The resulting Born scalar functions are
\ba
&&T_1^{\text{Born}}\left(\nu,\vec{q}\right)=-2i\left[\frac{Q^4\left(F_1^p+F_2^p\right)\left(F_1^Z+F_2^Z\right)}{Q^4-4M^2\nu^2}-F_1^pF_1^Z\right]
\nn\\
&&T_2^{\text{Born}}\left(\nu,\vec{q}\right)=\frac{-8iM\nu Q^2}{Q^4-4M^2\nu^2}\left(F_1^pF_1^Z+\frac{Q^2}{4M^2}F_2^pF_2^Z\right).
\nn\\
\ea
Expanding around $Q\simeq 0$ yields
\ba
\label{eq:T2delta}
&&T_1^{\text{Born}}\left(i\nu_E,\vec{q}\right)=2iF_1^p(0)F_1^Z(0)+O(Q^2)
\nn\\
&&T_2^{\text{Born}}\left(i\nu_E,\vec{q}\right)=4\pi \nu_E F_1^p(0)F_1^Z(0)\delta(\nu_E)+O(Q^2).
\nn\\
\ea
The contribution to $\square_{\gamma Z}^{V,\text{Wick},Q<\epsilon}(0)$ from $T_1^{\text{Born}}$ vanishes,
while the $\delta$-function in $T_2^{\text{Born}}$ yields a finite result.
Substituting Eq.~(\ref{eq:T2delta}) into Eq.~(\ref{eq:wickall}) gives
\ba
\label{eq:col_E=0}
		\square_{\gamma Z}^{V,\text{Wick},Q<\epsilon}(0)&=&-\frac{8\alpha_{\text{em}} g_A^e F_1^p(0)F_1^Z(0)}{\pi}\lim_{\epsilon\to 0}\lim_{E\to 0_{+}}\int_{0}^{\epsilon}dQ
		\nn\\
		&&\left(\frac{1}{E}+\frac{Q^2-4E^2}{4E^2Q}\ln\left|\frac{Q-2E}{Q+2E}\right|\right)
		\nn\\
		&=&-4\pi\alpha_{\text{em}} g_A^eF_1^{p}(0)F_1^Z(0)
		\nn\\
		&=&\pi\alpha_{\text{em}} Q_W^{p,\text{LO}}.
\ea

\subsection{Noise reduction for $\square_{\gamma Z}^V(E)$}

The vector $\gamma Z$ box, $\square_{\gamma Z}^{V}(E)$, is governed by the vector–vector Compton amplitude.
Both $J_\mu^{em}$ and $J_\mu^Z$ can be decomposed into isovector and isoscalar components,
$J_\mu^{I=1}=\frac{1}{2}(\bar{u}\gamma_\mu u-\bar{d}\gamma_\mu d)$
 and $J_\mu^{I=0}=\frac{1}{6}(\bar{u}\gamma_\mu u+\bar{d}\gamma_\mu d)$. Accordingly, the hadronic tensor separates into three channels, i.e. $\mathcal{H}_{\mu\nu}^{I,I'}(t_E,\vec{x})$
 with $\{I,I'\}=\{0,0\},\{0,1\},\{1,1\}$.
 Gauge invariance implies that, in the limit $(\nu,\vec{q})\to(0,\vec{0})$ and for sufficiently large $t_s$ such that contributions from all excited intermediate states except the nucleon vanishes,
\be
 \label{eq:SD_relation}
T_{00}^{\{I,I'\},\text{SD}}(0,\vec{0};t_s)=Mt_s,\quad \sum_i T_{ii}^{\{I,I'\},\text{SD}}(0,\vec{0};t_s)=\frac{3}{2},
\ee
for all three isospin channels. Here $T_{\mu\nu}^{\{I,I'\},\text{SD}}$ are defined by inserting $\mathcal{H}_{\mu\nu}^{I,I'}(t_E,\vec{x})$ into Eq.~(\ref{eq:T_SD}). The above relations were verified in our previous work~\cite{Fu:2022fgh,Wang:2023omf,Fu:2024gxq}.

\begin{figure}[htbp]
	\centering
	\includegraphics[scale=0.35]{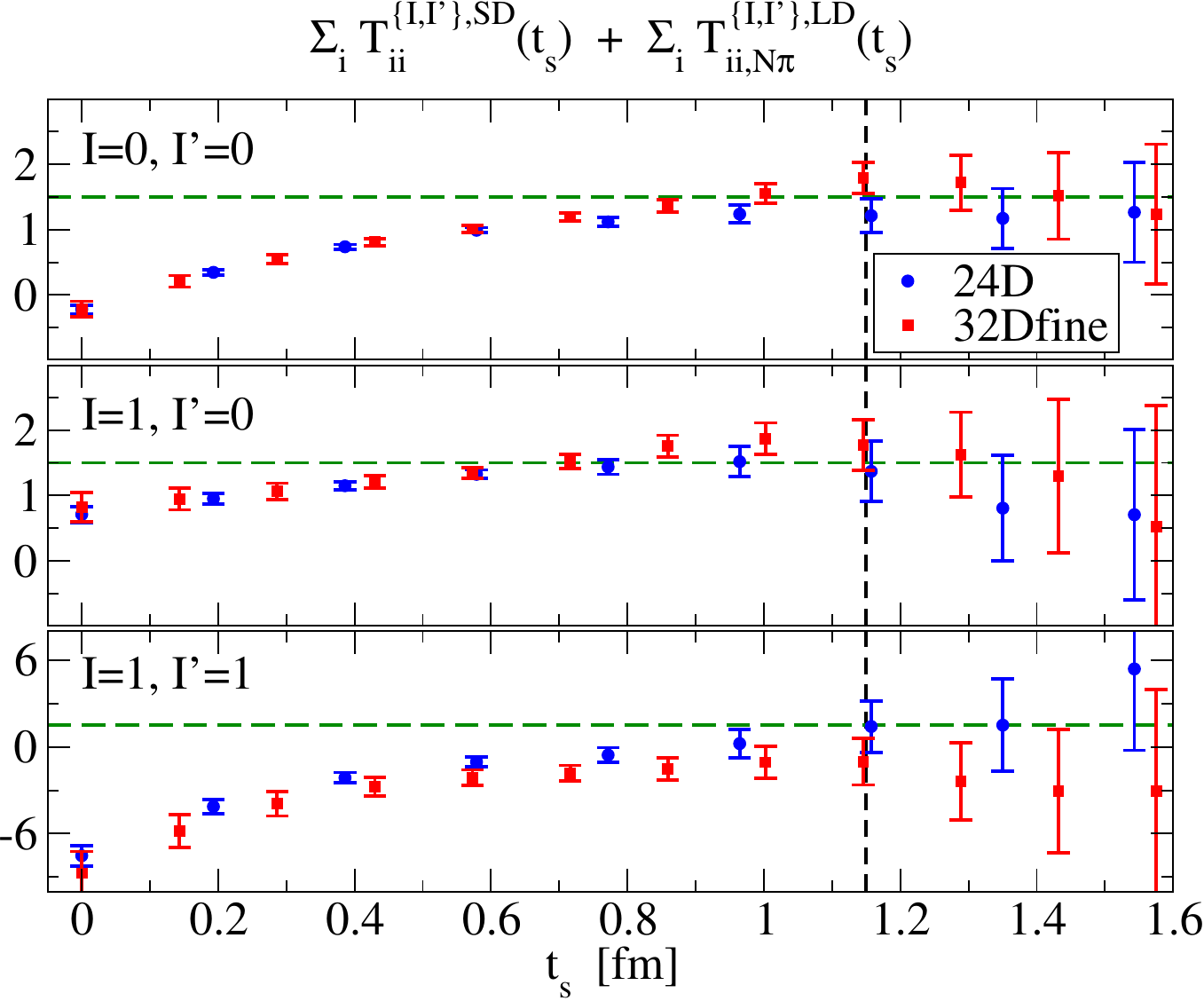}	
	\caption{Verification of Eq.~(\ref{eq:SD_relation_Npi}). For all three isospin $\{I,I'\}$ channels, the relations in (\ref{eq:SD_relation_Npi}) are satisfied for 
	$t_s>1.15$ fm on both 24D and 32Dfine ensembles.}
	\label{fig:check_Hii}
\end{figure}

In the present calculation, we also determine the $N\pi$ intermediate-state contribution, which permits a more stringent test of Eq.~(\ref{eq:SD_relation}) 
by adding the residual long-distance $N\pi$ contribution for $t>t_s$.
At $(\nu,\vec q)=(0,\vec 0)$, parity forbids $N\pi$ contributions to $T_{00}$, so the check focuses on $T_{ii}$. The long-distance $N\pi$ contribution is
\ba
T_{ii,N\pi}^{\{I,I'\},\text{LD}}(0,\vec{0};t_s)&=&2\sum_n\langle p|J_i^{I}|N\pi,n\rangle \langle N\pi,n|J_i^{I'}|p\rangle
\nn\\
&&\hspace{0.5cm}\times\frac{e^{-(E_{N\pi,n}-M)t_s}}{E_{N\pi,n}-M}
\ea
where $|N\pi,n\rangle$ denotes the low-lying center-of-mass $N\pi$ eigenstates, with $n$ labeling the energy level and the isospin channel ($I=1/2$ or $3/2$).
Using $N\pi$ states with lowest four energy eigenvalues as input, we find that for $t_s \gtrsim$ 1.15 fm
\be
\label{eq:SD_relation_Npi}
\sum_i T_{ii}^{\{I,I'\},\text{SD}}(0,\vec{0};t_s)+\sum_i T_{ii,N\pi}^{\{I,I'\},\text{LD}}(0,\vec{0};t_s)=\frac{3}{2}
\ee
The results are displayed in Fig.~\ref{fig:check_Hii}. 

The short-distance contribution $T_{\mu\nu}^{\{I,I'\},\text{SD}}(\nu,\vec{q};t_s)$ can be evaluated either directly from Eq.~(\ref{eq:T_SD}) or through
\ba
&&T_{00}^{\{I,I'\},\text{SD}}(\nu,\vec{q};t_s)
=Mt_s
\nn\\
&&\hspace{0.2cm}+\int_{-t_s}^{t_s}dt_E\int d^3\vec{x}\,(e^{\nu t_E}e^{-i\vec{q}\cdot\vec{x}}-1)\mathcal{H}_{00}^{I,I'}(t_E,\vec{x})
\nn\\
&&\sum_iT_{ii}^{\{I,I'\},\text{SD}}(\nu,\vec{q};t_s)
=\frac{3}{2}-\sum_i T_{ii,N\pi}^{\{I,I'\},\text{LD}}(0,\vec{0};t_s)
\nn\\
&&\hspace{0.2cm}+\int_{-t_s}^{t_s}dt_E\int d^3\vec{x}\,(e^{\nu t_E}e^{-i\vec{q}\cdot\vec{x}}-1)\sum_i\mathcal{H}_{ii}^{I,I'}(t_E,\vec{x}).
\nn\\
\ea
We further reduce statistical noise by forming a weighted average of the two determinations, using their statistical uncertainties as inverse-variance weights.
This averaging is valid only for $t_s>1.15$ fm, where the two determinations agree.
The resulting noise reduction is particularly effective for the determination of the residue contribution $\square_{\gamma Z}^{V,\mathrm{res}}$.

\subsection{Euclidean time dependence of the vector residue contribution}

\begin{figure}[h]
\centering
\includegraphics[width=0.45\textwidth]{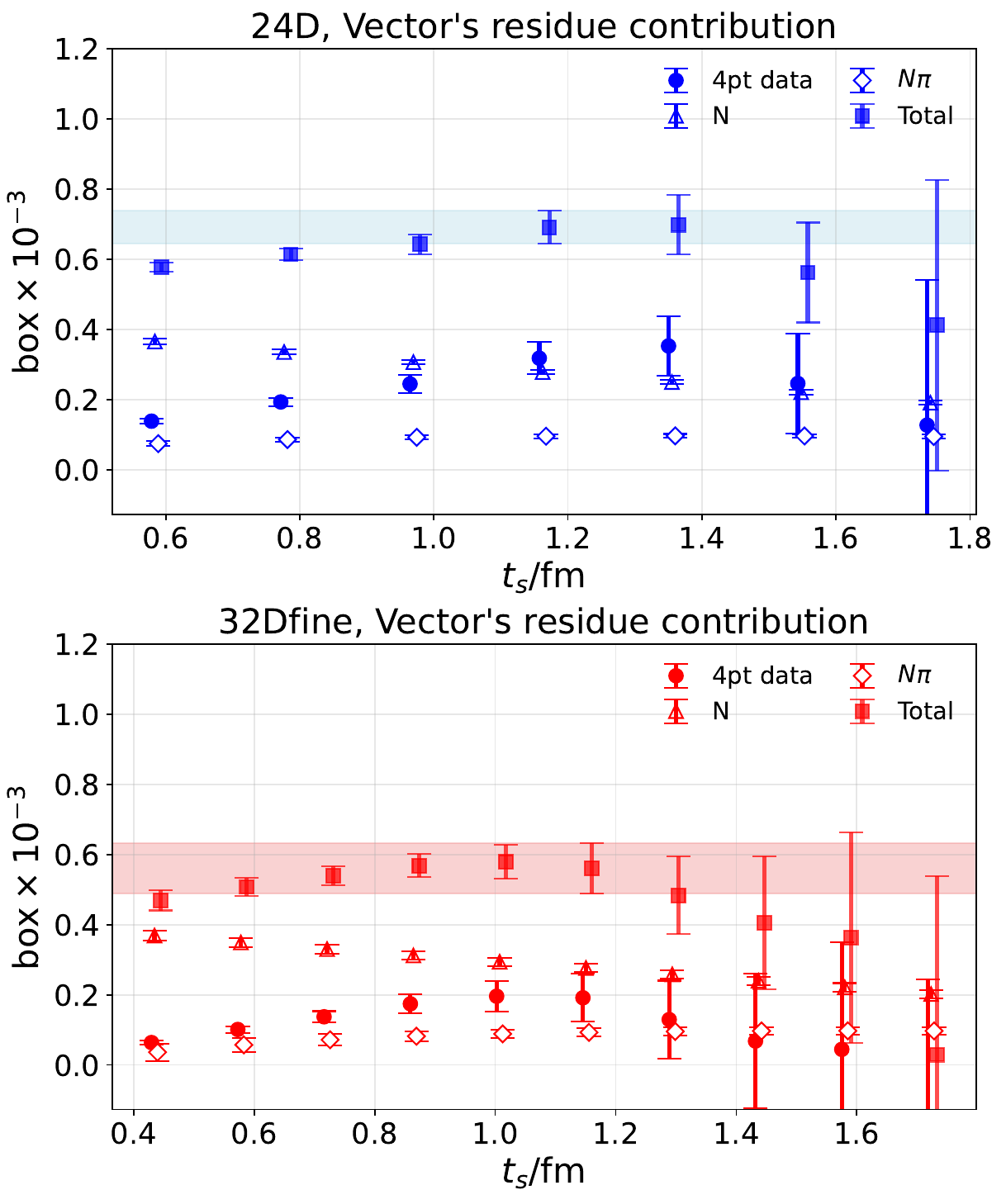}
\caption{Time-integrated vector residue contributions at $E=0.155$~GeV. We show the 4pt data, $N$ contribution, $N\pi$ contribution, and Total. The colored horizontal band marks the Total datapoint nearest $t\approx 1.15$~fm in each ensemble.}
\label{fig:pert_int}
\end{figure}

In Fig.~\ref{fig:pert_int} we present a decomposition of $\square_{\gamma Z}^V(E)$ into contributions from the four-point correlation functions (4pt), the reconstructed $N$ state ($N$), and the reconstructed $N\pi$ states ($N\pi$), as well as their combined result (total). This provides a diagnostic counterpart to Fig.~\ref{fig:vector} in the main text, where the result is presented in terms of short- and long-distance contributions. 
In the noise-reduction procedure, the short-distance contribution in Fig.~\ref{fig:vector} receives contributions from both the direct 4pt data and the reconstructed $N$ and $N\pi$ states. The two representations are fully consistent, differing only in how the contributions are organized. Fig.~\ref{fig:pert_int} shows that the total contribution leads to a stable plateau at $t_s\simeq 1.15$ fm.

\begin{figure}[h]
\centering
\includegraphics[width=0.45\textwidth]{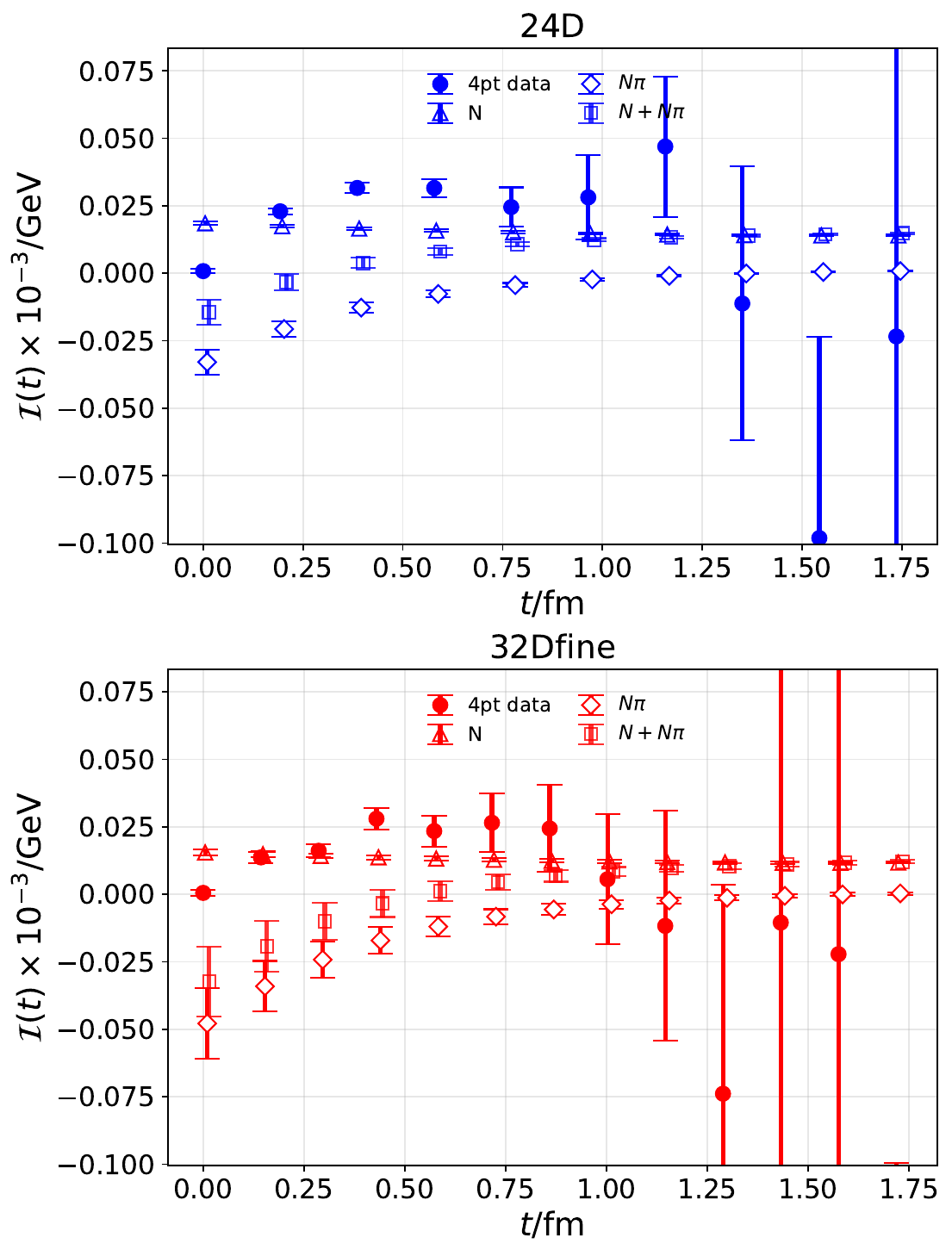}
\caption{Integrands for the vector residue contribution at $E=0.155$~GeV.}
\label{fig:pert_tdep_reorg}
\end{figure}

In Fig.~\ref{fig:pert_tdep_reorg} we show the corresponding integrands before time integration. At large temporal separations, the direct 4pt data become increasingly noisy, while the reconstructed $N$ and $N\pi$ contributions remain well behaved. We observe that the 4pt results are statistically consistent with the reconstructed $N+N\pi$ contributions within uncertainties at large time separation. The choice of $t \simeq 1.15$ fm used in the analysis corresponds to a conservative region where the large statistical uncertainty already covers the systematic difference between the direct and reconstructed contributions even at small $t$.

\begin{figure}[h]
\centering
\includegraphics[width=0.45\textwidth]{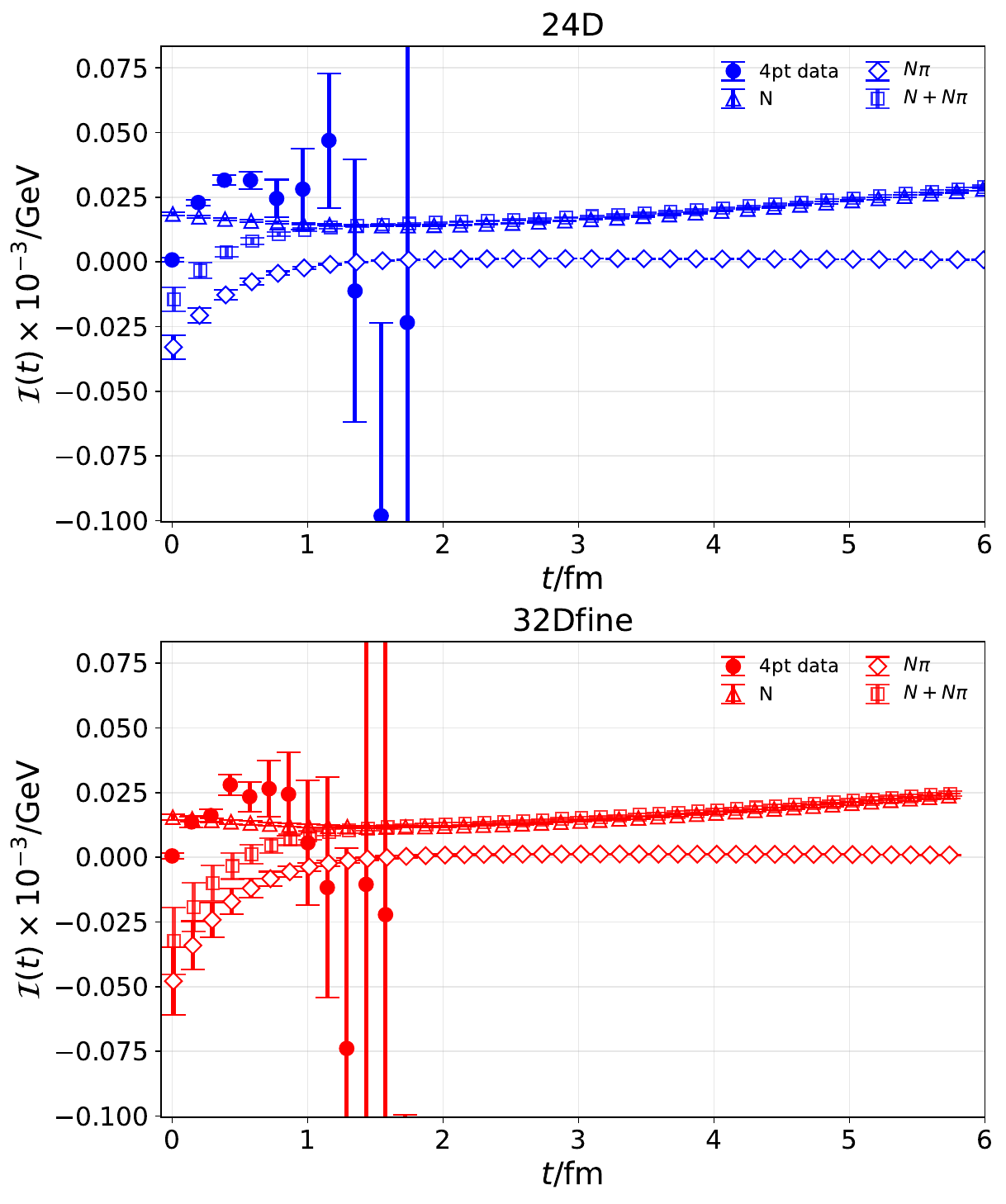}
\caption{Similar as Fig~\ref{fig:pert_tdep_reorg}, but with time window extended to 6 fm.}
\label{fig:pert_tdep_reorg_24d_tmax40}
\end{figure}

In Fig.~\ref{fig:pert_tdep_reorg_24d_tmax40} we illustrate the onset of the exponentially growing behavior at much larger time separations ($t \sim 6$ fm). Although such contributions would lead to a divergent long-distance integral if untreated, they are explicitly removed in our approach through the reconstruction of the $N$ and $N\pi$ intermediate states from three-point correlation functions.

\subsection{Calculation of $\square_{\gamma Z}^{A}(E)$}

\begin{figure}[htbp]
	\centering
	\includegraphics[scale=0.45]{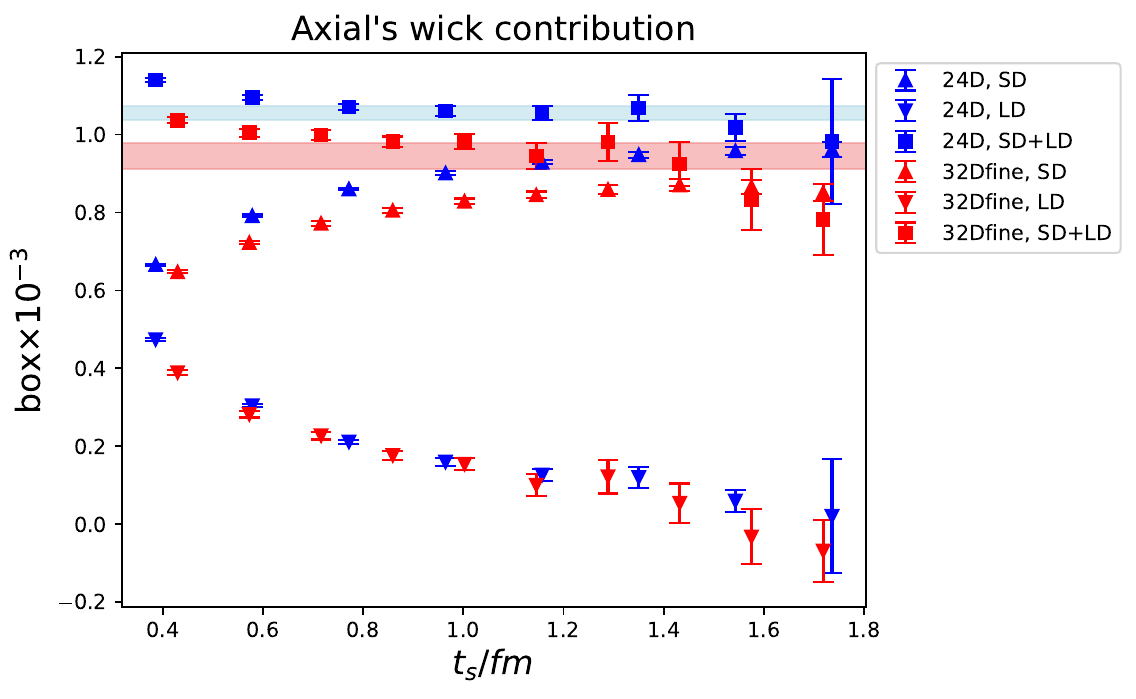}	
	\includegraphics[scale=0.45]{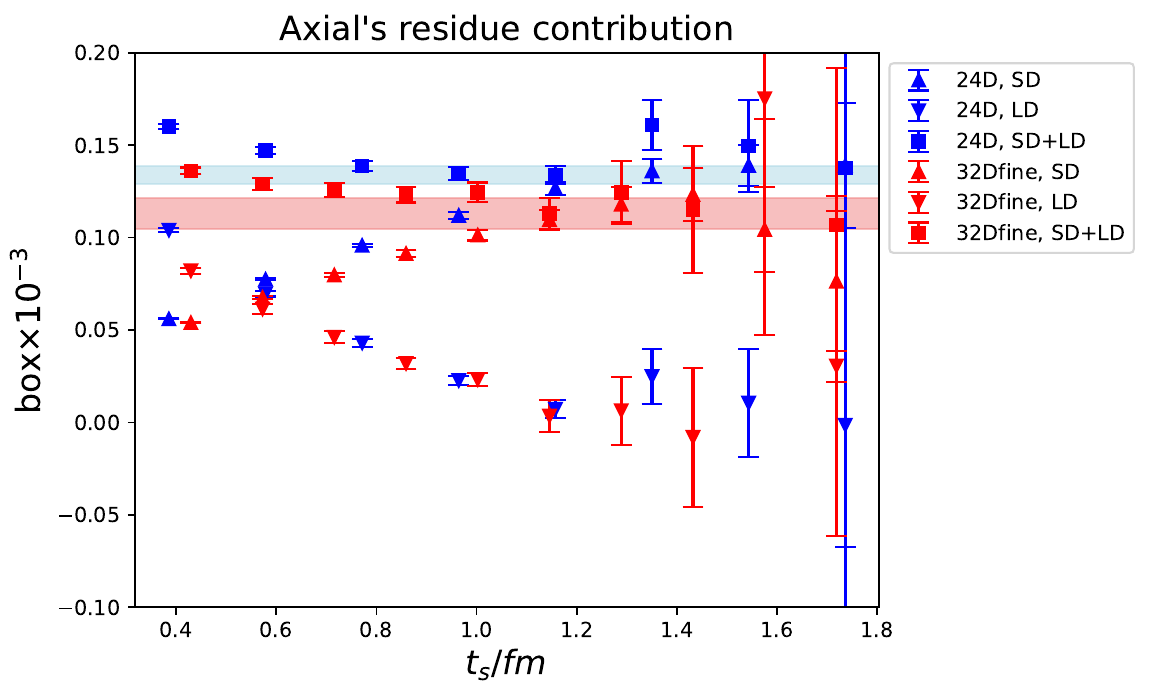}
	\caption{Numerical results for the Wick and residue contributions to $\square_{\gamma Z}^{A,Q^2<2~\mathrm{GeV}^2}(E)$ at a beam energy of $E=155$ MeV. The short-distance, the long-distance and the total contributions are indicated by upward triangles, downward triangles and squares, respectively.}
	\label{fig:axial_vector}
\end{figure}

The axial-vector $\gamma Z$ box contribution is separated into nonperturbative and perturbative components. First, we present the nonperturbative part determined from lattice QCD.
The lattice contribution is further decomposed into
short- and long-distance contributions by splitting the Euclidean time at $t_s$, denoted SD and LD, respectively. The LD part is reconstructed using IVR under the assumption of nucleon-state dominance. The residual $N\pi$ contribution is estimated using chiral perturbation theory ($\chi$PT)~\cite{Erler:2019rmr} and found to be $\sim 10^{-6}$ in $\square_{\gamma Z}^{A,Q^2<2~\mathrm{GeV}^2}(E)$ - three orders of magnitude 
below the total - and is therefore neglected. Figure~\ref{fig:axial_vector} shows the resulting SD, LD, and total (SD+LD) contributions.

The perturbative component $\square_{\gamma Z}^{A,Q^2>2~\mathrm{GeV}^2}(E)$ is computed using the operator product expansion; details are provided in Ref.~\cite{Erler:2019rmr}. This contribution exhibits only mild dependence on the beam energy $E$. At $E=155$ MeV, we obtain
\begin{equation}
	\square_{\gamma Z}^{A,Q^2>2\text{GeV}^2}(E)=3.029(6)_{\text{HO}}(6)_{\text{HT}}\times 10^{-3},
\end{equation}
where the first uncertainty reflects higher-order perturbative truncation, determined from the difference between 4-loop and 3-loop results, 
and the second captures higher-twist truncation, estimated from diagrams involving two currents on different quark lines, similar to the strategy for $\gamma W$ box in Ref.~\cite{Feng:2020zdc}.

\subsection{Calculation of Sachs form factors $G_{E(M)}^{\text{em(Z)}}(Q^2)$}

\begin{figure}[htbp]
	\centering
	\includegraphics[scale=0.6]{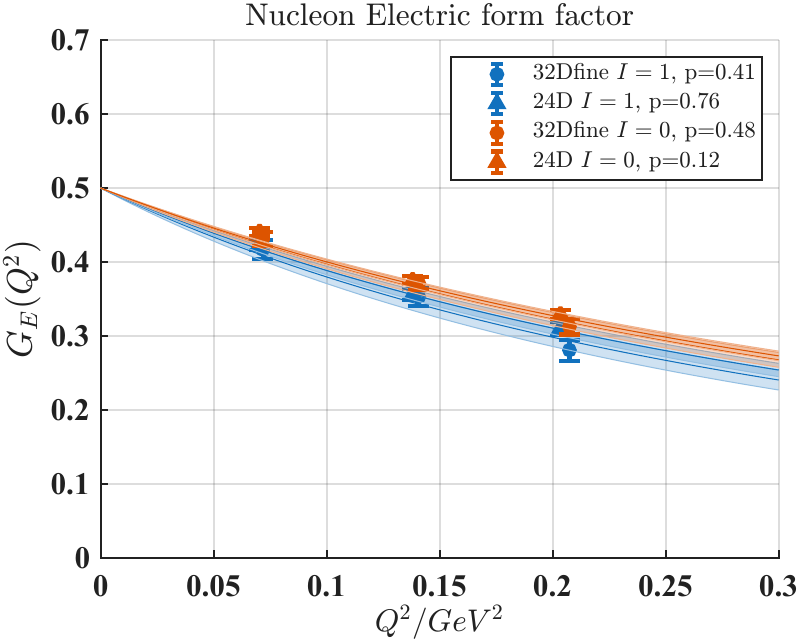}
	\includegraphics[scale=0.6]{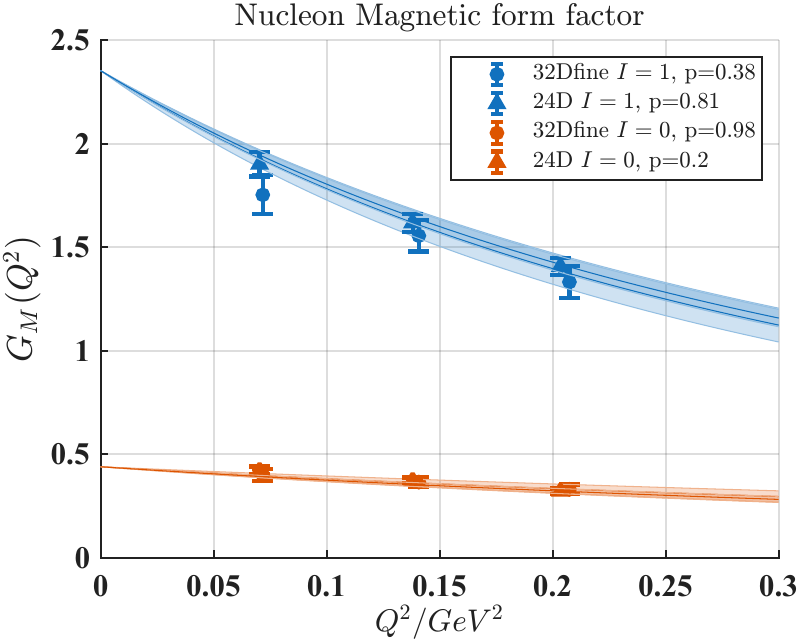}
	\caption{Numerical results for the $I=0$ and $I=1$ Sach form factors defined in Eq.~(\ref{eq:Sach_def}) together with a dipole fit.}
	\label{Fig:Sach_FF}
\end{figure}

The Sachs form factors are determined from nucleon three-point correlation functions
\begin{equation}
	\langle O_p(\vec{p},t_{\text{snk}}) J_{\mu}(t) \overline{O}_p(\vec{0},t_{\text{src}})\rangle,
\end{equation}
where the spatial momentum projectors fix the proton momenta to
$\vec{p}$ and $\vec{0}$ at the sink and source, respectively.
To isolate the ground-state matrix element, we construct appropriate ratios of three-point to two-point functions and perform a summation over the operator insertion time
$t_{\text{src}}<t<t_{\text{snk}}$, which suppresses excited-state contamination. At sufficiently large source-sink separation, 
the resulting matrix elements determine the isovector ($I=1$) and isoscalar ($I=0$) Sachs form factors
\ba
\label{eq:Sach_def}
		&&\bra{p(\vec{p})}J_{\mu}^{I=1}\ket{p(\vec{0})}
\nn\\
		&&\hspace{1cm}=\overline{u}(\vec{p})\left(\gamma_{\mu}F_1^{V}(Q^2)+\frac{i\sigma_{\mu\nu}q^{\nu}}{2M}F_2^{V}(Q^2)\right)u(0)
\nn\\
		&&\bra{p(\vec{p})}J_{\mu}^{I=0}\ket{p(\vec{0})}
\nn\\
		&&\hspace{1cm}=\overline{u}(\vec{p})\left(\gamma_{\mu}F_1^{S}(Q^2)+\frac{i\sigma_{\mu\nu}q^{\nu}}{2M}F_2^{S}(Q^2)\right)u(0)
\nn\\
		&&G_E^{V(S)}(Q^2)=F_1^{V(S)}(Q^2)+\frac{q^2}{4M^2}F_2^{V(S)}(Q^2)
\nn\\
		&&G_M^{V(S)}(Q^2)=F_1^{V(S)}(Q^2)+F_2^{V(S)}(Q^2)
\ea
where $Q^2$ denotes the Euclidean momentum transfer.
Our lattice results for $G_{E(M)}^{V(S)}(Q^2)$ are shown in Fig.~\ref{Fig:Sach_FF}. The data are fitted using a dipole parametrization
\ba
&&G_E^{V(S)}(Q^2)=\frac{1/2}{\left(1+Q^2/b_E^{V(S)}\right)^2},
\nn\\ 
&&G_M^{V(S)}(Q^2)=\frac{\left(\mu_p\mp \mu_n\right)/2}{\left(1+Q^2/b_M^{V(S)}\right)^2},
\ea
where $b_{E(M)}^{V(S)}$ is a fit parameter and $\mu_{p(n)}$ denotes the proton (neutron) magnetic moment fixed to PDG values~\cite{ParticleDataGroup:2024cfk}; the resulting curves are also displayed in Fig.~\ref{Fig:Sach_FF}.

Using the current definition in Eq.~(\ref{eq:currents}), the electromagnetic and neutral-weak-current Sachs form factors are obtained as
\ba
		G_{E(M)}^{\text{em}}(Q^2)&=&G_{E(M)}^{V}(Q^2)+G_{E(M)}^{S}(Q^2)
		\nn\\
		G_{E(M)}^{\text{Z}}(Q^2)&=&(g_V^{u}-g_V^d)G_{E(M)}^{V}(Q^2)
		\nn\\
		&&+3(g_V^{u}+g_V^d)G_{E(M)}^{S}(Q^2).
\ea
From the momentum dependence of the form factors, we extract the proton and neutron electric and magnetic radii.
The results obtained on the 24D and 32Dfine ensembles, together with the continuum-extrapolated ones, are summarized in Table~\ref{tab:zero_mom} and are in overall agreement with the PDG values.

\begin{table}[htbp]
	\centering
	\begin{tabular}{ccccc}
		\hline
		\hline
		& 24D & 32Dfine & Cont. & PDG \\
		\hline
		$\sqrt{r_E^2}\big|_{p}$ & $0.767(19)$ & $0.791(29)$ & $0.820(69)$ & $0.8409(4)$\\
		$\sqrt{|r_E^2|}\big|_{n}$ & $0.147(27)$ & $0.188(34)$ & $0.235(83)$ & $0.339(25)$	\\
		$\sqrt{r_M^2}\big|_{p}$ & $0.786(25)$ & $0.792(49)$ & $0.80(11)$ & $0.851(26)$	\\
		$\sqrt{r_M^2}\big|_{n}$ & $0.859(27)$ & $0.908(53)$ & $0.97(12)$ & $0.864(9)$\\
		\hline
	\end{tabular}%
 \caption{Numerical results for the electric and magnetic radii. Results are shown for both ensembles, together with the continuum-extrapolated values; the PDG values are listed in the final column for comparison. All quantities are given in $\text{fm}$.}
\label{tab:zero_mom}
\end{table}

We compute the Born contribution $\square_{\gamma Z}^{V,\mathrm{Born}}(E)$ using the lattice form-factor fits shown in Fig.~\ref{Fig:Sach_FF}, 
and compared it with phenomenological expectations in Table~\ref{tab:vector_decomp}. The table also shows the corresponding total and non-Born contributions.

\begin{table}[htbp]
\centering
\begin{tabular}{lcccc}
\hline
Component & 32Dfine & 24D & Continuum & Phenomenology \\
\hline
Born & 1.77(10) & 1.91(06) & 1.60(24) & 1.87 \\
Total & 2.28(13) & 2.63(09) & 1.84(31) & 2.94(18) \\
Non-Born & 0.51(13) & 0.73(09) & 0.23(31) & 1.07(18) \\
\hline
\end{tabular}
\caption{Vector contribution decomposition into elastic (Born) and non-elastic (non-Born) contributions. All values are in units of $10^{-3}$.}
\label{tab:vector_decomp}
\end{table}

We find that the elastic (Born) contribution is in good agreement with the phenomenological estimate within uncertainties. In contrast, the non-Born contribution is about $2.3\sigma$ smaller than the phenomenological expectation for both ensembles and after continuum extrapolation.
This comparison provides a useful benchmark of our treatment of the elastic pole, and shows that the difference with phenomenology resides predominantly in the non-Born contribution. Understanding the origin of this difference warrants further study.

\subsection{Long-distance reconstruction}

To address both Wick and residue contributions, we consider the general form
\be
\label{eq:FV_wick}
I(i\nu_E,\vec{q})=\int_{-T/2}^{T/2} dt_E\, e^{-i\nu_E t_E} \int_{-L/2}^{L/2} d^3\vec{x}\,e^{-i\vec{q}\cdot\vec{x}}H_L(t_E,\vec{x}),
\ee
where $\nu_E$ is real for the Fourier transform and $\nu_E=i\nu_e$ for the Laplace transform.
The hadronic function $H_L(t_E,\vec x)$ is defined in finite volume, while the momentum $\vec q$ is taken as a continuous variable corresponding to the infinite volume.
This setup follows the spirit of the $\text{QED}_\infty$ approach, in which the QED part is evaluated in infinite volume.
Since physical observables are ultimately defined in the infinite-volume limit, one may always replace the plane-wave factor by its angular average, $e^{-i\vec{q}\cdot\vec{x}}\to j_0(|\vec{q}||\vec{x}|)$, 
by assuming that the hadronic function is spatially 
SO(3) symmetric.
However, such a mixed treatment - combining an infinite-volume weight with a finite-volume hadronic function - does not guarantee exponential suppression of finite-volume
 effects if poles arise from low-lying intermediate states.

To illustrate this, we assume the large-$|t_E|$ behavior ($|t_E|>t_s$) of $H_L$ is dominated by the $N$ and $N\pi$ states
\ba
&&H_L(t_E,\vec{x})
\nn\\
&\approx& \frac{1}{L^3}\sum_{\vec{k}}\rho_L^{(N)}(\vec{k},E_N)e^{-(E_N(\vec{k})-M)|t_E|}e^{i\vec{k}\cdot\vec{x}}
\nn\\
&+&\frac{1}{L^3}\sum_{\vec{k},\Lambda,n}\rho_{L}^{(N\pi)}(\vec{k},E_{\Lambda,n})e^{-(E_{\Lambda,n}(\vec{k})-M)|t_E|}e^{i\vec{k}\cdot\vec{x}},
\nn\\
\ea
where $E_N(\vec{k})=\sqrt{M^2+\vec{k}^2}$ is the energy for a single nucleon while $E_{\Lambda,n}$ denotes the discrete $N\pi$ energy levels
classified by the cubic-group irreducible representation $\Lambda$ and level index $n$.
We decompose the integral as $I=I^{\text{SD}}+I^{\text{LD}}$:
\ba
I^{\text{SD}}&=&\int_{-t_s}^{t_s} dt_E\, e^{-i\nu_E t_E} \int_{-L/2}^{L/2} d^3\vec{x}\,j_0(|\vec{q}||\vec{x}|)H_L(t_E,\vec{x})
\nn\\
I^{\text{LD}}&=&\int_{t_s}^{T/2} dt_E\, e^{-i\nu_E t_E} \int_{-L/2}^{L/2} d^3\vec{x}\,j_0(|\vec{q}||\vec{x}|)H_L(t_E,\vec{x})
\nn\\
&+&\int_{-T/2}^{-t_s} dt_E\, e^{-i\nu_E t_E} \int_{-L/2}^{L/2} d^3\vec{x}\,j_0(|\vec{q}||\vec{x}|)H_L(t_E,\vec{x})
\nn\\
&=&\frac{1}{L^3}\sum_{\vec{k}}\rho_L^{(N)}(\vec{k},E_N)K_L(|\vec{q}|,\vec{k})
\nn\\ 
&&\hspace{1cm}\times\left(f(T/2,E_N(\vec{k}))-f(t_s,E_N(\vec{k}))\right)
\nn\\
&+&\frac{1}{L^3}\sum_{\vec{k},\Lambda,n}\rho_{L}^{(N\pi)}(\vec{k},E_{\Lambda,n})K_L(|\vec{q}|,\vec{k})
\nn\\
&&\hspace{1cm}\times\left(f(T/2,E_{\Lambda,n}(\vec{k}))-f(t_s,E_{\Lambda,n}(\vec{k}))\right)
\nn\\
\ea 
with
\ba 
&&K_L(|\vec{q}|,\vec{k})=\int_{-L/2}^{L/2} d^3\vec{x}\,j_0(|\vec{q}||\vec{x}|)e^{-i\vec{k}\cdot\vec{x}},
\nn\\
&&f(t,E_N(\vec{k}))=\frac{e^{(-i\nu_E-E_N(\vec{k})+M)t}}{-i\nu_E-E_N(\vec{k})+M}+\{\nu_E\to-\nu_E\}.
\nn\\
\ea
Terms involving $T$ vanish at large $T$ or can be removed explicitly.

For the nucleon contribution, a pole appears in $\frac{1}{-i\nu_E - E_N(\vec k)+M}$ when $\nu_E\to0$ (Fourier case) or $\nu_E=i\nu_e$ (Laplace case), leading to power-law-suppressed finite-volume effects.
To further reduce it to the exponentially suppressed level, we adopt the IVR and define $I^{\text{LD},(N)}$ by replacing $\vec{k}$ with $\vec{q}$
\ba
I^{\text{LD},(N)}&=&\frac{1}{L^3}\sum_{\vec{k}}\rho_L^{(N)}(\vec{k},E_N)K_L(|\vec{q}|,\vec{k})
\nn\\
&\times&\left(
\frac{-e^{(-i\nu_E-E_N(\vec{k})+M)t_s}}{-i\nu_E-E_N(\vec{q})+M}+\{\nu_E\to-\nu_E\}\right).
\nn\\
\ea
In the infinite-volume limit, $K_L(|\vec{q}|,\vec{k})\to K_\infty(|\vec{q}|,\vec{k})=\frac{2\pi^2}{|\vec{q}|^2}\delta(|\vec{k}|-|\vec{q}|)$, confirming that the replacement $\vec{k}\to\vec{q}$ 
recovers the correct infinite-volume result while eliminating the $\vec{k}$-dependent pole.
According to the Poisson summation formula, $I^{\mathrm{LD},(N)}$ thus exhibits only exponentially-suppressed finite-volume corrections.
Integrating over $\vec{q}$ yields
\ba
\square^{\text{LD},(N)}_{\text{Wick}}&=&\mathcal{PV} \int\frac{d^3\vec{q}}{(2\pi)^3} \int\frac{d\nu_E}{2\pi}V(i\nu_E,\vec{q})I^{\text{LD},(N)}(i\nu_E,\vec{q})
\nn\\
\square^{\text{LD},(N)}_{\text{res}}&=&\mathcal{PV} \int\frac{d^3\vec{q}}{(2\pi)^3} \operatorname{Res}[V(q)]I^{\text{LD},(N)}(\nu_e,\vec{q})
\ea
where $V(i\nu_E,\vec{q})$ and $\operatorname{Res}[V(q)]$ originates from Eq.~(\ref{eq:Wick_residue}).

For the $N\pi$ contribution, the situation is more involved. The discrete energies $E_{\Lambda,n}$ are determined by L\"uscher's quantization condition and depend on the input 
$N\pi$ scattering phase shift, preventing a simple replacement $\vec{k}\to\vec{q}$. Moreover, in the Laplace transform, the 
$N\pi$ states with energies $E_{\Lambda,n}$ satisfying
\be
\nu_e-E_{\Lambda,n}+M\ge 0
\ee
generate exponentially growing factors. Since $\nu_e=E-|\vec{l}-\vec{q}|$ attains its maximal value at $\vec{q}\approx\vec{l}$,
the condition
\be
\label{eq:exp_growing_cond}
E-E_{\Lambda,n}+M\ge 0
\ee
identifies the states that can induce exponentially enhanced contamination. For the lattice volumes used in this study, only the
$N\pi$ ground state with $|\vec{k}|=0$ and $E_{\Lambda,n}\approx M+M_\pi$ satisfies this condition, while
higher excitations with larger $|\vec{k}|$ or $n$ yield exponentially suppressed factors. 
Nonetheless, the low-lying modes may still contribute to the long-distance part, albeit exponentially damped.
To quantify this effect, we write
\be
\label{eq:LD_Npi}
\square^{\text{LD},(N\pi)}_{\text{res}}=\frac{1}{L^3}\sum_{\vec{k},\Lambda,n}\rho_{L}^{(N\pi)}(\vec{k},E_{\Lambda,n})g_{\text{res}}^{\text{LD}}(\vec{k},E_{\Lambda,n})
\ee 
with the factor $g_{\text{res}}^{\text{LD}}(\vec{k},E_{\Lambda,n})$ defined as
\ba
g_{\text{res}}^{\text{LD}}(\vec{k},E_{\Lambda,n})&=&\mathcal{PV}\int\frac{d^3\vec{q}}{(2\pi)^3}\operatorname{Res}[V(q)]K_L(|\vec{q}|,\vec{k})
\nn\\
&& \times\left(-f(t_s,E_{\Lambda,n})+\{\nu_E\to-\nu_E\}\right)\Big|_{\nu_E=-i\nu_e}
\ea
For large $E_{\Lambda,n}$ or $\vec{k}$, $g_{\text{res}}^{\text{LD}}(\vec{k},E_{\Lambda,n})$ is exponentially suppressed by $f(t_s,E_{\Lambda,n})$ or kinematically suppressed
by $K_L(|\vec{q}|,\vec{k})$, leaving only low-lying modes as significant contributors.
To estimate the typical size of $g_{\text{res}}^{\text{LD}}(\vec{k},E_{\Lambda,n})$, we neglect the $N\pi$ interactions and approximate
\ba
\label{eq:simplification}
&\sum_{\vec{k},\Lambda,n}\to\sum_{\vec{k},\vec{p}_N}&,
\nn\\
&E_{\Lambda,n}(\vec{k})\to E_{N\pi}(\vec{k},\vec{p}_N)=E_N(\vec{p}_N)+E_\pi(\vec{k}-\vec{p}_N)&
\nn\\
&g_{\text{res}}^{\text{LD}}(\vec{k},E_{\Lambda,n})\to g_{\text{res}}^{\text{LD}}(\vec{k},E_{N\pi}(\vec{k},\vec{p}_N))&
\ea
The dominant terms of $g_{\text{res}}^{\text{LD}}(\vec{k},E_{N\pi}(\vec{k},\vec{p}_N))$ at $E=155$ MeV and $t_s=1.15$ fm are listed in Table~\ref{tab:gres}. 
Using these as benchmarks, we compute the lattice spectra $E_{\Lambda,n}$ and corresponding spectral weight $\rho_{L}^{(N\pi)}(\vec{k},E_{\Lambda,n})$ for the relevant low-lying modes.
Note that Eq.~(\ref{eq:simplification}) serves only to gauge the relative magnitude of $g_{\text{res}}^{\text{LD}}(\vec{k},E_{N\pi}(\vec{k},\vec{p}_N))$; in the actual reconstruction we employ lattice-determined 
$E_{\Lambda,n}$ and $\rho_{L}^{(N\pi)}(\vec{k},E_{\Lambda,n})$ directly. The full $N\pi$ contribution to the residue term is then obtained from Eq.~(\ref{eq:LD_Npi}) with $g_{\text{res}}^{\text{LD}}(\vec{k},E_{\Lambda,n})\to g_{\text{res}}(\vec{k},E_{\Lambda,n})$, where
\ba
g_{\text{res}}(\vec{k},E_{\Lambda,n})&=&\mathcal{PV}\int\frac{d^3\vec{q}}{(2\pi)^3}\operatorname{Res}[V(q)]K_L(|\vec{q}|,\vec{k})
\nn\\
&& \times\left(-f(0,E_{\Lambda,n})+\{\nu_E\to-\nu_E\}\right)\Big|_{\nu_E=-i\nu_e}
\nn\\
&=&\mathcal{PV}\int\frac{d^3\vec{q}}{(2\pi)^3}\operatorname{Res}[V(q)]K_L(|\vec{q}|,\vec{k})
\nn\\
&&\hspace{1cm}\times\frac{2(E_{\Lambda,n}(\vec{k})-M)}{(E_{\Lambda,n}(\vec{k})-M)^2-\nu_e^2}
\ea

\begin{table}[htbp]
	\centering
	\begin{tabular}{cc}
		\hline
		\hline
		Momentum modes & $g_{\text{res}}^{\text{LD}}(\vec{k},E_{N\pi}(\vec{k},\vec{p}_N))$ \\
		\hline
		$\vec{k}=\vec{0}$, $|\vec{p}_N|=0$ & $3.29\times10^{-3}$\\
		$\vec{k}=\vec{0}$, $|\vec{p}_N|=\frac{2\pi}{L}$ & $3.01\times10^{-4}$	\\
		$\vec{k}=\vec{0}$, $|\vec{p}_N|=\sqrt{2}\frac{2\pi}{L}$ & $9.29\times10^{-5}$	\\
		$\vec{k}=\vec{0}$, $|\vec{p}_N|=\sqrt{3}\frac{2\pi}{L}$ & $3.75\times10^{-5}$	\\
		$\vec{k}=\vec{p}_N$, $|\vec{p}_N|=\frac{2\pi}{L}$ & $2.73\times10^{-4}$	\\
		$\vec{k}=\vec{p}_N$, $|\vec{p}_N|=\sqrt{2}\frac{2\pi}{L}$ & $-5.62\times10^{-5}$	\\
		$\vec{k}=\vec{p}_N$, $|\vec{p}_N|=\sqrt{3}\frac{2\pi}{L}$ & $-3.48\times10^{-5}$	\\
		\hline
	\end{tabular}%
 \caption{The dominant terms of $g_{\text{res}}^{\text{LD}}(\vec{k},E_{N\pi}(\vec{k},\vec{p}_N))$ at $E=155$ MeV and $t_s=1.15$ fm.}
\label{tab:gres}
\end{table}

\subsection{Sensitivity to variations of $Q_{\mathrm{cut}}^2$}

We use to $Q_{\rm cut}^2=1,2,3$ GeV$^2$ to assess the sensitivity of the results to variations of $Q_{\rm cut}^2$.

\begin{table}[htbp]
\centering
\begin{tabular}{cccc}
\hline
$Q_{\mathrm{cut}}^2$  & Lattice & Perturbative & Total \\
(GeV$^2$)  & Cont. extrap. & (4-loop) & \\
\hline
1  & 0.732(90) & $3.195(12)_{\mathrm{HO}}(20)_{\mathrm{HT}}$ & 3.926(93) \\
2  & 0.898(91) & $3.029(6)_{\mathrm{HO}}(6)_{\mathrm{HT}}$ & 3.927(91) \\
3  & 0.997(91) & $2.921(4)_{\mathrm{HO}}(2)_{\mathrm{HT}}$ & 3.918(91) \\
\hline
\end{tabular}
\caption{Axial box contribution for different $Q_{\rm cut}^2$. All values are in units of $10^{-3}$. For the perturbative column, the first error is labeled $_{\mathrm{HO}}$ (3-loop vs 4-loop) and the second is labeled $_{\mathrm{HT}}$ (higher-twist).}
\label{tab:Qcut}
\end{table}

As shown in Table~\ref{tab:Qcut}, varying $Q_{\rm cut}^2$ in the range $1~{\rm GeV}^2 \le Q_{\rm cut}^2 \le 3~{\rm GeV}^2$ leads to a 
shift in the total result at the level of $\sim 0.01 \times 10^{-3}$, which is much smaller than the quoted total uncertainty. 
This indicates that the matching between the lattice and perturbative contributions 
is stable at the current level of precision.

In this work, we take this variation as an estimate of the residual matching uncertainty, 
which is numerically negligible compared to the total error and does not affect the final result.

\subsection{Finite-volume effects}
As discussed above, for the nucleon contribution, the IVR method suppresses the finite-volume effects to an exponentially small level. 
We now turn to the finite-volume corrections associated with the $N\pi$ intermediate state.
The total $N\pi$ contribution to the residue term can be written as
\ba
\label{eq:Npi_res}
&&\square^{(N\pi)}_{\text{res}}
\nn\\
&=&\frac{1}{L^3}\sum_{\vec{k},\Lambda,n}\rho_{L}^{(N\pi)}(\vec{k},E_{\Lambda,n})g_{\text{res}}(\vec{k},E_{\Lambda,n})
\nn\\
&=&\frac{1}{L^3}\sum_{\vec{k},\Lambda,n}\mathcal{PV}\int\frac{d^3\vec{q}}{(2\pi)^3}
\frac{2(E_{\Lambda,n}(\vec{k})-M)C(\vec{q},\vec{k},E_{\Lambda,n})}{(E_{\Lambda,n}(\vec{k})-M)^2-\nu_e^2}
\nn\\
\ea
where
\be
C(\vec{q},\vec{k},E_{\Lambda,n})=\rho_{L}^{(N\pi)}(\vec{k},E_{\Lambda,n})\operatorname{Res}[V(q)]K_L(|\vec{q}|,\vec{k})
\ee
Since $\nu_e \ge 0$ and $E_{\Lambda,n} > M$, the integrand contains a single singularity at
\be
E_{\Lambda,n}(\vec{k})-M-\nu_e=0
\ee

\subsubsection{Step 1: Principal-value analysis of the momentum integral}

We next introduce the variable transformation $(|\vec{q}|,\cos\theta)\to(|\vec{q}|,E_e)$, with
\be
E_e=|\vec{q}-\vec{l}|=\sqrt{|\vec{q}|^2+E^2-2\cos\theta |\vec{q}|E},
\ee
for which
\be
d|\vec{q}|d\cos\theta= \frac{E_e}{|\vec{q}|E} d|\vec{q}|dE_e,
\ee
and the kinematic constraint $E \ge |\vec{q}-\vec{l}|$ implies
\be
|\vec{q}|\le 2E,\quad \frac{|\vec{q}|}{2E} \le \cos\theta\le 1
\ee
We can rewrite $C(\vec{q},\vec{k},E_{\Lambda,n})$ as $C(|\vec{q}|,E_e,\vec{k},E_{\Lambda,n})$.
The variable substitution leads to
\ba
\square^{(N\pi)}_{\text{res}}&=&\frac{1}{4\pi^2 E}\frac{1}{L^3}\sum_{\vec{k},\Lambda,n}\mathcal{PV}\int_0^{2E}  d|\vec{q}|\int_{|E-|\vec{q}||}^{E} dE_e
\nn\\
&&\hspace{0.5cm}\frac{2(E_{\Lambda,n}(\vec{k})-M)E_e |\vec{q}|  C(|\vec{q}|,E_e,\vec{k},E_{\Lambda,n})}{(E_{\Lambda,n}(\vec{k})-M)^2-(E-E_e)^2}
\nn\\
&=&\frac{1}{4\pi^2 E}\frac{1}{L^3}\sum_{\vec{k},\Lambda,n}\mathcal{PV}\int_0^{E}  d|\vec{q}|\int_{E-|\vec{q}|}^{E} dE_e
\nn\\
&&\hspace{0.5cm}\frac{2(E_{\Lambda,n}(\vec{k})-M)\bar{C}(|\vec{q}|,E_e,\vec{k},E_{\Lambda,n})}{(E_{\Lambda,n}(\vec{k})-M)^2-(E-E_e)^2},
\ea
where
\ba
&&\bar{C}(|\vec{q}|,E_e,\vec{k},E_{\Lambda,n})
\nn\\
&=&E_e \left[|\vec{q}|C(|\vec{q}|,E_e,\vec{k},E_{\Lambda,n})+\{|\vec{q}|\to 2E-|\vec{q}|\}\right].
\ea
The integrand exhibits a pole at
\be
\label{eq:E_e0}
E_{e,0}=E- (E_{\Lambda,n}-M)
\ee
We decompose the integral as $\square^{(N\pi),(1)}_{\text{res}}+\square^{(N\pi),(2)}_{\text{res}}$ by splitting
\be
\label{eq:splitting}
\bar{C}=\left(\bar{C}-\bar{C}\left|_{E_e=E_{e,0}}\right)+\bar{C}\right|_{E_e=E_{e,0}}.
\ee
Although the integral limits satisfy $E-|\vec{q}|<E_e<E$, the function $\bar{C}(|\vec{q}|,E_e,\vec{k},E_{\Lambda,n})$ is analytic in $E_e$; hence, it is legitimate to choose the subtraction point 
$E=E_{e,0}$ even if it lies outside the physical region.
The first term in Eq.~(\ref{eq:splitting}) removes the pole in the denominator and thus yields only exponentially suppressed corrections.
The second term contains the power-law finite-volume contribution
\ba
\square^{(N\pi),(2)}_{\text{res}}&=&\frac{1}{4\pi^2 E}\frac{1}{L^3}\sum_{\vec{k},\Lambda,n}\mathcal{PV}\int_0^{E} d|\vec{q}| \,\bar{C}(|\vec{q}|,E_{e,0},\vec{k},E_{\Lambda,n})
\nn\\
&&\int_{E-|\vec{q}|}^{E} dE_e\,
\frac{2(E_{\Lambda,n}(\vec{k})-M)}{(E_{\Lambda,n}(\vec{k})-M)^2-(E-E_e)^2}
\nn\\
&=&\frac{1}{4\pi^2 E}\frac{1}{L^3}\sum_{\vec{k},\Lambda,n}\mathcal{PV}\int_0^{E} d|\vec{q}| \,\bar{C}(|\vec{q}|,E_{e,0},\vec{k},E_{\Lambda,n})
\nn\\
&&\hspace{1cm}\ln\left(\left|\frac{E_{\Lambda,n}-M+|\vec{q}|}{E_{\Lambda,n}-M-|\vec{q}|}\right|\right).
\ea
A singularity appears at
\be
\label{eq:q_0}
|\vec{q}|=q_0\equiv E_{\Lambda,n}-M
\ee
provided $E_{\Lambda,n}-M<E$. Hence, the power-law finite-volume effect resides solely in
\be
\square^{(N\pi),(3)}_{\text{res}}=\frac{1}{4\pi^2 E}\frac{1}{L^3}\sum_{\vec{k},\Lambda,n} \bar{C}(q_0,E_{e,0},\vec{k},E_{\Lambda,n}) I(E_{\Lambda,n})
\ee
where
\ba
I(E_{\Lambda,n})&=&\mathcal{PV}\int_0^{E} d|\vec{q}| \,\ln\left(\left|\frac{E_{\Lambda,n}-M+|\vec{q}|}{E_{\Lambda,n}-M-|\vec{q}|}\right|\right)
\nn\\
&=&(E_{\Lambda,n}-M+E)\ln(E_{\Lambda,n}-M+E)
\nn\\
&+&(E_{\Lambda,n}-M-E)\ln |E_{\Lambda,n}-M-E|
\nn\\
&-&2(E_{\Lambda,n}-M)\ln(E_{\Lambda,n}-M).
\ea

\subsubsection{Step 2: Analysis of nonanalytic behavior}

The nonanalytic behavior originates from a removable singular point located at
\be
\label{eq:E_Lambda}
E_{\Lambda,n}=M+E
\ee
Combining Eqs.~(\ref{eq:E_e0}), (\ref{eq:q_0}) and (\ref{eq:E_Lambda}), the corresponding kinematics read
\be
\label{eq:kinematics}
\vec{q}=\vec{l},\quad |\vec{q}|=q_0=E,\quad \theta=0,\quad E_{e,0}=0,\quad E_{\Lambda,n}=M+E.
\ee

According to the theorem of Ref.~\cite{Tuo:2024bhm}, 
the magnitude of finite-volume (FV) effects depends on the smoothness of the integrand: if the integrand is continuously differentiable up to order $N$, the FV correction scales as $O(1/L^{N+1})$, with $L$ the spatial extent of the lattice. We therefore examine the smoothness of
$\bar{C}(q_0,E_{e,0},\vec{k},E_{\Lambda,n})$ near the kinematic point specified in Eq.~(\ref{eq:kinematics}).

The quantity $\bar{C}(q_0,E_{e,0},\vec{k},E_{\Lambda,n})$ depends on the FV spectral density $\rho_L^{N\pi}(\vec{k},E_{\Lambda,n})$, which is defined at discrete energies $E_{\Lambda,n}$.
Introducing Lellouch-L\"uscher factor $f_{LL}(\vec{k},E_{\Lambda,n})$, the finite- and infinite-volume spectral weights are related by
\be
\rho_\infty(\vec{k},E_{\Lambda,n})=f_{LL}(\vec{k},E_{\Lambda,n})\rho_L(\vec{k},E_{\Lambda,n}).
\ee
The function $\rho_\infty$ generally depends on both $\vec{k}$ and $E_{\Lambda,n}$, as well as on the relative orientation of $\vec{k}$ and the nucleon momentum $\vec{p}$; 
this angular dependence can be expanded in spherical harmonics and will not be discussed further.

Suppose there exists a smooth function $\bar{C}^0(q_0,E_{e,0},\vec{k},E_{\Lambda,n})$ satisfying
\be
\bar{C}(E,0,{\vec{k}},M+E)=\bar{C}^0(E,0,\vec{k},M+E).
\ee
Then $\bar{C}$ can be decomposed as
\be
\bar{C}=\bar{C}^0+\delta \bar{C}
\ee
where $\bar{C}^0$ captures the leading FV contribution, while $\delta\bar{C}$ contributes only to higher-order corrections.

At threshold, $E=M_\pi$, the construction of $\bar{C}^0$ is straightforward: only the lowest $N\pi$ state with $\vec{k}=\vec{0}$ satisfies $E_{\Lambda,n}=M+M_\pi$. In this case,
\ba
&&\bar{C}^0(q_0,E_{e,0},\vec{k},E_{\Lambda,n})
\nn\\
&=&2E f_{LL}^{-1}(\vec{k},E_{\Lambda,n})\rho_{\infty}^{(N\pi)}(\vec{0},M+M_\pi)
\nn\\
&&\times (E_{e,0}\operatorname{Res}[V(q)])\big|_{|\vec{q}|=E,E_{e,0}=0} K_L(E,\vec{k})
\nn\\
&\equiv&f_{LL}^{-1}(\vec{k},E_{\Lambda,n}) \mathcal{F}(E)K_L(E,\vec{k})
\ea
where $\rho_\infty^{N\pi}$ is fixed at $\vec{k}=\vec{0}$ and $E_{\Lambda,n}=M+M_\pi$.

The leading power-law FV contribution is then
\be
\square^{(N\pi),(4)}_{\text{res}}=\frac{1}{4\pi^2 E}\frac{1}{L^3}\sum_{\vec{k}\,\Lambda,n}\bar{C}^0(q_0,E_{e,0},\vec{k},E_{\Lambda,n}) I(E_{\Lambda,n}).
\ee
The corresponding infinite-volume expression reads
\ba
\square^{(N\pi),(4)}_{\text{res},\infty}&=&\frac{1}{4\pi^2 E}\int\frac{d^3\vec{k}}{(2\pi)^3}
\nn\\
&&\mathcal{PV}\sumint_{\alpha_{N\pi}}\mathcal{F}(E)K_\infty(E,\vec{}k) I(E_{N\pi}),
\ea
where $\sumint_{\alpha_{N\pi}}$ denotes the integral over infinite-volume $N\pi$ states
$|\alpha_{N\pi}(\vec{k},E_{N\pi})\rangle$.

The FV correction can be separated as
\be
\square^{(N\pi),(4)}_{\text{res}}-\square^{(N\pi),(4)}_{\text{res},\infty}=\Delta_{\text{FV},1}+\Delta_{\text{FV},2}
\ee
with
\ba
\Delta_{\text{FV},1}&=&\frac{\mathcal{F}(E)}{4\pi^2 E}\frac{1}{L^3}\sum_{\vec{k}}K_L(E,\vec{k})
\nn\\
&&\left(\sum_{\Lambda,n} f_{LL}^{-1}I(E_{\Lambda,n}) -\mathcal{PV}\sumint_{\alpha_{N\pi}} I(E_{N\pi})\right)  
\nn\\
\Delta_{\text{FV},2}&=&\frac{\mathcal{F}(E)}{4\pi^2 E}\left(\frac{1}{L^3}\sum_{\vec{k}} K_L -\int\frac{d^3\vec{k}}{(2\pi)^3} K_\infty\right)
\nn\\
&&\mathcal{PV}\sumint_{\alpha_{N\pi}} I(E_{N\pi})
\ea
Because $\Delta_{\text{FV},2}$ involves an additional integral over the continuum variable $E_{N\pi}$, 
it is parametrically suppressed relative to $\Delta_{\text{FV},1}$ and represents a higher-order FV effect.

For $E>M_\pi$, multiple $(\vec{k},E_{N\pi})$ combinations satisfy $E_{N\pi}=M+E$. The spectral weight $\rho_\infty(\vec{k},E_{N\pi}=M+E)$ imposes the constraint
\be
\label{eq:constraint}
\vec{k}^2\le (E+M)^2- (M+M_\pi)^2.
\ee
This condition is consistent with Eq.~(\ref{eq:exp_growing_cond}), used to identify the states responsible for exponentially growing contributions.
For $E=155$ MeV, this yields $|\vec{k}|\lesssim180$ MeV (using $M=940$ MeV and $M_\pi=140$ MeV). Hence, we approximate
\be
\label{eq:approx}
\rho_\infty(\vec{k},E+M)\approx \rho_\infty(\vec{0},M+M_\pi),\quad \text{for $|\vec{k}|\le 180$ MeV}
\ee
which enables an efficient numerical evaluation of $\Delta_{\text{FV},1}$ and $\Delta_{\text{FV},2}$.
The systematic uncertainty associated with this approximation is assessed in $\chi$PT, as discussed below.

\subsubsection{Step 3: Numerical estimation}

To estimate the FV corrections, $\Delta_{\text{FV},1}$ and $\Delta_{\text{FV},2}$, we neglect $N\pi$ rescattering effects for simplicity.
Under this approximation, the Lellouch–Lüscher factor $f_{LL}^{-1}$ reduces to a normalization accounting only for the state degeneracy,
\be
\sum_{\Lambda,n}f_{LL}^{-1} \to \frac{1}{L^3}\sum_{\vec{p}},\quad \sumint_{\alpha_{N\pi}}\to \int \frac{d^3\vec{p}}{(2\pi)^3}
\ee
where $\vec{p}$ denotes the nucleon momentum and the $N\pi$ energy satisfies
$E_{N\pi}=\sqrt{M^2+\vec{p}^2}+\sqrt{M_\pi^2+(\vec{k}-\vec{p})^2}$.

With this replacement, the FV corrections take the approximate form
\ba
\Delta_{\text{FV},1}&=&\frac{\mathcal{F}(E)}{4\pi^2 E}\frac{1}{L^3}\sum_{\vec{k}}K_L(E,\vec{k})
\nn\\
&&\left(\frac{1}{L^3}\sum_{\vec{p}}  -\mathcal{PV}\int\frac{d^3\vec{p}}{(2\pi)^3}\right)  I(E_{N\pi})  
\nn\\
\Delta_{\text{FV},2}&=&\frac{\mathcal{F}(E)}{4\pi^2 E}\left(\frac{1}{L^3}\sum_{\vec{k}} K_L -\int\frac{d^3\vec{k}}{(2\pi)^3} K_\infty\right)
\nn\\
&&\mathcal{PV}\int\frac{d^3\vec{p}}{(2\pi)^3} I(E_{N\pi})
\ea

The ratios of the FV corrections to the total vector $\gamma Z$ box contribution, $\square_{\gamma Z}^V$, are summarized in Table~\ref{tab:FV}.

\begin{table}[htbp]
	\centering
	\begin{tabular}{c|cc|cc}
		\hline
		\hline
		 & \multicolumn{2}{c|}{$E=M_\pi$} & \multicolumn{2}{c}{$E=155$ MeV} \\
		\hline
		& 24D & 32Dfine & 24D & 32Dfine \\
		\hline
		$\Delta_{\text{FV},1}/\square_{\gamma Z}^V$  & 3.8(3)\% & 4.6(7)\% & 4.1(3)\% & 5.1(8)\% \\
		$\Delta_{\text{FV},2}/\square_{\gamma Z}^V$  &   0.14(1)\% & 0.17(3)\% & 0.15(1)\%  &  0.19(3)\%	\\
		\hline
	\end{tabular}%
 \caption{Estimation of leading-order FV effects from $N\pi$ states.}
\label{tab:FV}
\end{table}

In our previous work~\cite{Wang:2023omf}, we have verified that the lattice results for $\rho_\infty(\vec{0},M+M_\pi)$ are consistent with $\chi$PT within uncertainties.
We therefore use $\chi$PT to assess the systematic effects associated with the approximation in Eq.~(\ref{eq:approx}).
In general, $\rho_\infty(\vec{k},M+E)$ decreases with increasing $|\vec{k}|$ and $M+E$, implying that fixing it to $\rho_\infty(\vec{0},M+M_\pi)$ slightly overestimates the FV effects.
By varying $\rho_\infty(\vec{k},M+E)$ within the range constrained by Eq.~(\ref{eq:constraint}), we find that $\Delta_{\text{FV},1}$ and $\Delta_{\text{FV},2}$ are reduced by approximately $19\sim 33\%$. 
In all cases, the leading FV corrections from $N\pi$ states remain small compared with the overall statistical uncertainty and are included as a source of systematic error.

\subsection{Excited-state effects}

\begin{figure}[htbp]
	\centering
	\includegraphics[width=0.45\textwidth]{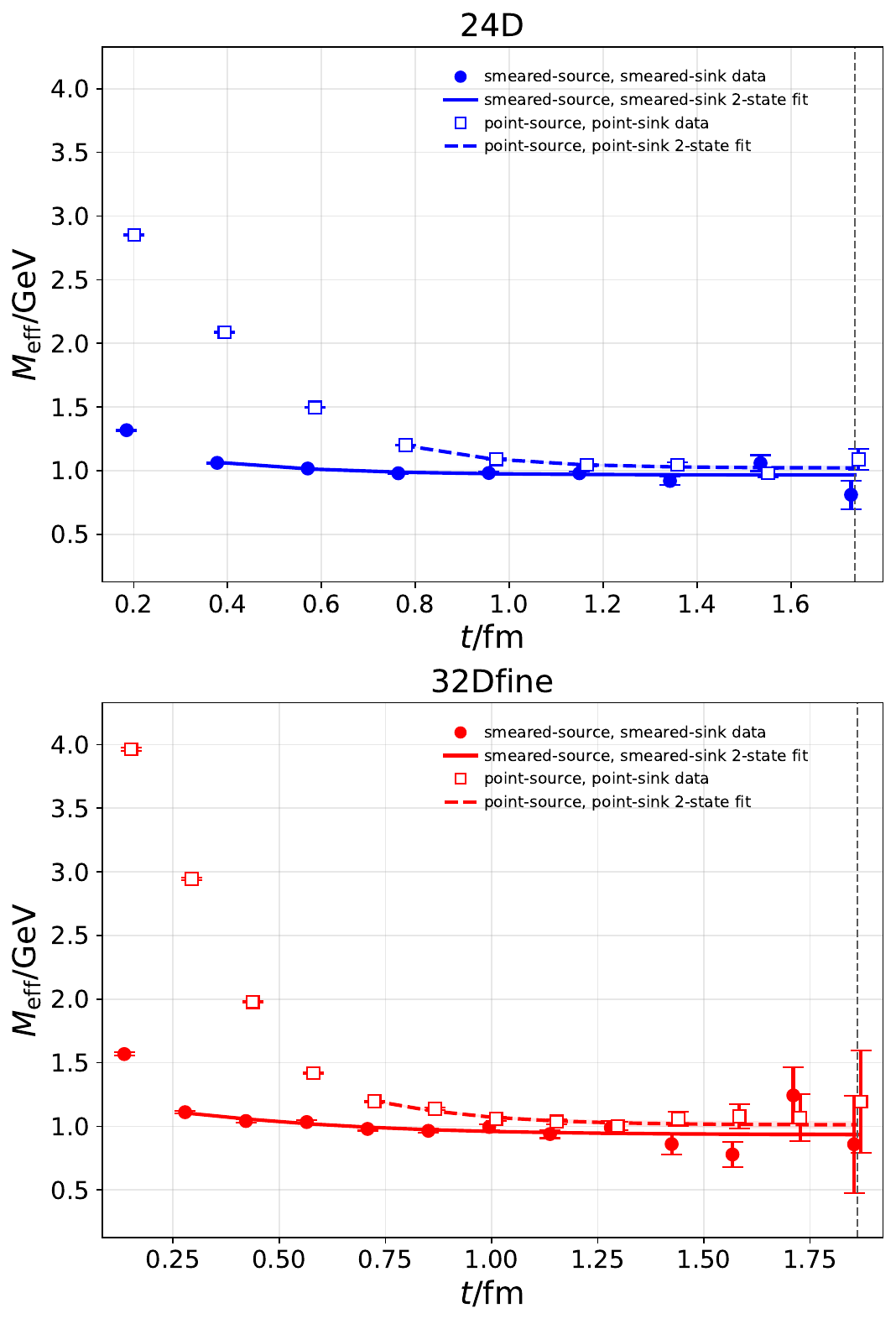}
	\caption{Comparison of the effective mass $M_{\mathrm{eff}}(t)=\log(C(t)/C(t+1))/a$ for the proton point-source-point-sink and smeared-source-smeared-sink correlators. The solid and dashed curves show the corresponding two-state fit. For each case, the displayed time range matches the fit window.}
\label{fig:psrc_ssrc_fitwindow}
\end{figure}

 \begin{figure}[htb]
\centering
\includegraphics[width=0.48\textwidth,angle=0]{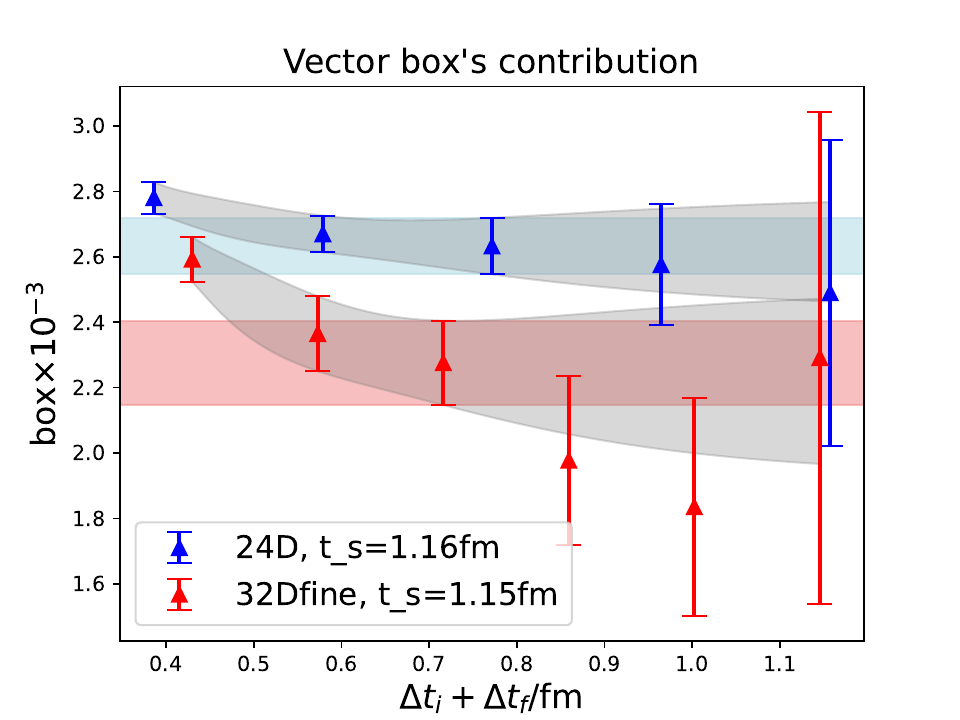}
\includegraphics[width=0.48\textwidth,angle=0]{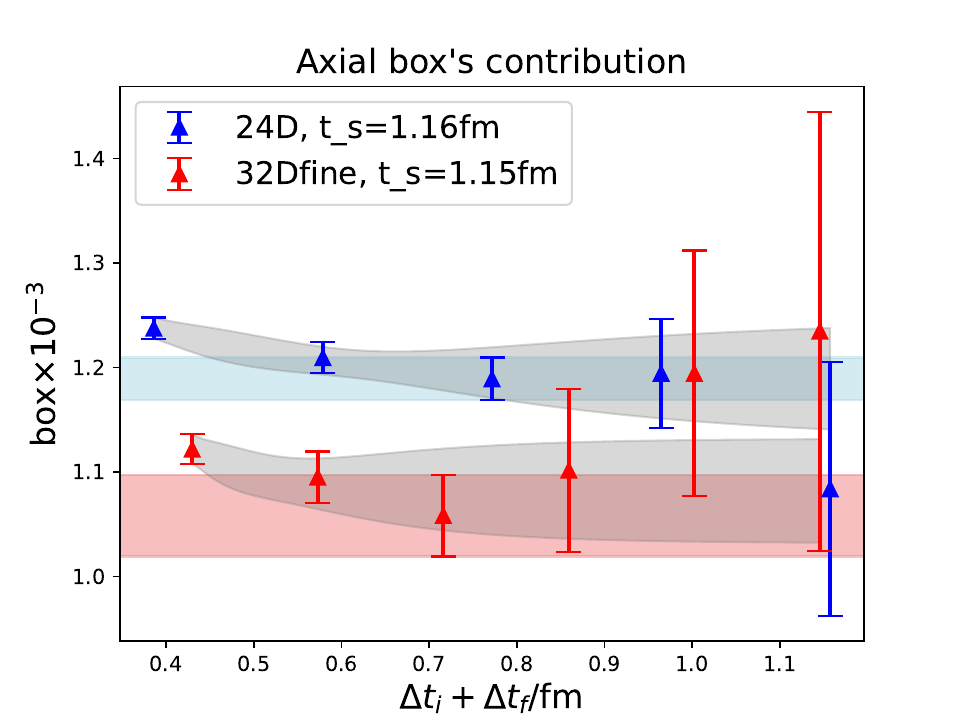}
\caption{The vector and axial-vector $\gamma Z$ box as a function of $\Delta t_i+\Delta t_f$. We employ a two-state fit to examine the excited-state contaminations.}
\label{fig:t_dep}
\end{figure}

Smeared sources and sinks are employed throughout to improve the signal-to-noise
ratio and suppress excited-state contamination from the nucleon interpolators. To quantify
the impact of smearing, we compare in Fig.~\ref{fig:psrc_ssrc_fitwindow}
the nucleon eﬀective masses obtained using
smeared and point sources, together with two-state fits. A clear reduction of excited-state
contamination is observed when using smeared operators. The corresponding fit results
for the ground-state and first excited-state energies are summarized in Table~\ref{tab:psrc_ssrc_twostate}, showing improved ground-state saturation for smeared sources.

\begin{table}[htbp]
\centering
\begin{tabular}{llcccc}
\hline
Ensemble & Source  & $E_0$ & $E_1$ & $\Delta E$  & $\chi^2/\text{d.o.f.}$ \\
\hline
24D & Smeared  & 0.966(8) & 1.79(18) & 0.83(17) & 1.01 \\
24D & Point  & 1.020(19) & 2.10(20) & 1.08(18) & 0.39 \\
32Dfine & Smeared  & 0.934(23) & 1.53(16) & 0.60(14) & 0.94 \\
32Dfine & Point  & 1.01(3) & 1.93(24) & 0.92(21) & 0.35 \\
\hline
\end{tabular}
\caption{Two-state fit results for the nucleon point-source and smeared-source two-point correlation functions. We quote the fitted ground-state energy $E_0$, first excited-state energy $E_1$, the gap $\Delta E = E_1-E_0$, and the fit quality. The energies are shown in units of GeV.}
\label{tab:psrc_ssrc_twostate}
\end{table}

To quantify the residual excited-state contamination in the final observable, 
we plot the vector and axial-vector $\gamma Z$ box contributions at $E=155$ MeV 
as a function of $\Delta t_i+\Delta t_f$ in Fig.~\ref{fig:t_dep}. 
A plateau is observed starting at $\Delta t_i+\Delta t_f\simeq 0.75$ fm. 
We then perform two-state fits to estimate the residual effects. 
For fits without constraint on the energy gap, we obtain the shift between the fit and the result at $\Delta t_i+\Delta t_f\simeq 0.75$ fm, 
\ba
&&\text{Vector:}\quad \Delta^{\text{24D}}=-2(11)\times10^{-5},\,\, \Delta^{\text{32Dfine}}=-6(21)\times10^{-5}
\nn\\
&&\text{Axial:}\quad \Delta^{\text{24D}}=0(5)\times10^{-5},\,\, \Delta^{\text{32Dfine}}=2(4)\times10^{-5}.
\nn\\
\ea
These shifts are statistically consistent with zero. 

\begin{table}[htbp]
\centering
\begin{tabular}{llccc}
\hline
Channel & Ensemble & $\Delta E^{\mathrm{2pt}}_{\mathrm{smear}}$  & $\Delta E^{\mathrm{4pt}}$ & $\Delta_{\mathrm{fixed-ref}} \times 10^{5}$ \\
\hline
Vector & 24D & 0.83(17) & 1.12(2.02)  & $-4.4(4.5)$ \\
Vector & 32Dfine & 0.60(14)  & 1.30(1.75) & $-25.9(10.8)$ \\
Axial & 24D & 0.83(17) & 0.64(1.17)  & $+0.67(1.46)$ \\
Axial & 32Dfine & 0.60(14)  & 0.99(2.12)  & $+1.89(3.39)$ \\
\hline
\end{tabular}
\caption{Consistency check between the two-point and four-point excited-state gaps. The 4pt gap is extracted from the unconstrained $a+b e^{-\Delta E t}$ fit, while the constrained refit uses the smeared two-point gap. The last column gives the shift between the constrained-fit asymptote and the reference $\Delta t_i+\Delta t_f$ point, in units of $10^{-5}$.}
\label{tab:dtdep_gap_consistency}
\end{table}

The energy gaps extracted from the four-point functions are compared with those 
from smeared two-point functions in Table~\ref{tab:dtdep_gap_consistency}. 
The energy gaps extracted from the four-point functions do not show a statistically significant deviation from those obtained from the smeared two-point functions.

As a further cross-check, we perform constrained fits using the energy gaps 
from smeared two-point functions. The resulting shifts $\Delta_{\mathrm{fixed-ref}}$ 
are summarized in Table~\ref{tab:dtdep_gap_consistency}.

In this work, we conservatively take the larger of the shifts from 
unconstrained and constrained fits as the systematic uncertainty 
from residual excited-state effects, which is included in the final error budget.

\end{document}